\documentclass[reprint,superscriptaddress, amsmath,amssymb, aps, pra, ]{revtex4-2}
\usepackage[usenames, dvipsnames]{color}
\usepackage{comment}
\usepackage{hyperref}
\usepackage{subcaption}
\usepackage{float}
\usepackage{ragged2e}
\hypersetup{
    colorlinks=true,
    linkcolor=Blue,
    citecolor=Blue,
    filecolor=Blue,      
    urlcolor=Blue,
    pdftitle={Quasiparticle Properties Long Range Impurity}}
\usepackage{graphicx}
\usepackage{physics}

\def\betak{\beta_{\mathbf{k}}}
\def\ptot{\mathbf{p}_{\mathrm{tot}}}
\def\betaVec{\Vec{\beta}}
\def\Psir{\Psi(\mathbf{r})}
\def\Psirconj{\Psi^*(\mathbf{r})}
\def\UIBr{U_{\mathrm{IB}}(\mathbf{r})}
\def\delpsir{\delta \psi(\mathbf{r})}
\def\delpsirt{\delta \psi(\mathbf{r},t)}

\def\delpsirtconj{\delta \psi^*(\mathbf{r},t)}
\def\hkkPrime{h_{\mathbf{k}\mathbf{k'}}}
\def\delpsik{\delta \psi_{\mathbf{k}}}
\def\epsilonkr{\epsilon_{k_r}}
\def\Ekr{E_{k_r}}
\def\aIBinverse{a_{\mathrm{IB}}^{-1}}
\def\aPlusinverse{a_{+}^{-1}}

\def\gammakv{\hbar \mathbf{k} \cdot \mathbf{V}}
\def\gammakvSquare{(\hbar \mathbf{k} \cdot \mathbf{V})^2}
\newcommand{\delpsikk}[1]{\delta \psi_{\mathbf{#1}}}
\newcommand{\betakk}[1]{\beta_{\mathbf{#1}}}
\newcommand{\sumk}[1]{\sum_{\mathbf{#1}}}
\newcommand{\thetak}[1]{\theta_{#1}}
\newcommand{\UIBk}[1]{\mathcal{U}_{\mathrm{IB}}(\mathbf{#1})}
\newcommand{\kdotv}[1]{\mathbf{#1}\cdot \mathbf{V}}
\begin{document}

\title{Quasiparticle properties of long-range impurities in a Bose condensate}

\author{T. A. Yo\u{g}urt}
 \email{ayogurt@pks.mpg.de}
 \affiliation{%
Max Planck Institute for the Physics of Complex Systems, N\"othnitzer Str. 38, 01187 Dresden, Germany\\
}%
 \affiliation{%
Department of Physics, Middle East Technical University, Ankara, 06800, Turkey\\
}%

\author{Matthew T. Eiles}
\email{meiles@pks.mpg.de}
 \affiliation{%
Max Planck Institute for the Physics of Complex Systems, N\"othnitzer Str. 38, 01187 Dresden, Germany\\
}%
\date{\today}

\begin{abstract}
An impurity immersed in a Bose condensate can form a quasiparticle known as a Bose polaron. When the impurity-boson interaction is short-ranged, the quasiparticle properties can be characterized in terms of the impurity-boson scattering length $a_{\mathrm{IB}}$ and the condensate coherence length $\xi$, a universal description that remains valid irrespective of the bath density $n_0$.
Long-ranged interactions—such as provided by Rydberg or ionic impurities—introduce an effective interaction range $r_{\mathrm{eff}}$ as the third length scale.
These competing length scales raise the question of whether a universal description remains valid across different bath densities. In this study, we discuss the quasiparticle nature of long-range impurities and its dependence on the length scales $n_0^{-1/3}$, $r_\mathrm{eff}$, and $\xi$. We employ two complementary theories—the coherent state Ansatz and the perturbative Gross-Pitaevskii theory—which incorporate beyond-Fröhlich interactions. We derive an analytical expression for the beyond-Fr\"ohlich effective mass for a contact interaction and numerically compute the effective mass for long-range impurities. We argue that the coupling parameter $|a_{\mathrm{IB}}|n_0^{1/3}$ remains the principal parameter governing the properties of the polaron. For weak ($|a_\mathrm{IB}|n_0^{1/3}\ll 1$) and intermediate ($|a_\mathrm{IB}|n_0^{1/3}\simeq 1$) values of the coupling parameter, long-range impurities in a Bose condensate are well-described as quasiparticles with a finite quasiparticle weight and a well-defined effective mass. However, the quasiparticle weight becomes significantly suppressed as the effective impurity volume is occupied by an increasing number of bath particles ($r_{\mathrm{eff}}n_0^{1/3} \gg 1$). 

\end{abstract}

\maketitle

\section{\label{sec: Introduction} INTRODUCTION}

Atomic impurities embedded in weakly-interacting Bose gases \cite{2016_PRL_Cornell_Bose_Polaron_Experiment,2016_PRL_Bruun_Bose_Polaron_Experiment,2020_Science_Zwierlien_Bose_Polaron,2024_Arxiv_Cambridge_Bose_Polaron} provide a highly-controllable and versatile environment with which to study quasiparticle physics \cite{1948_Landau_Pekar_Effective_Mass,1954_Frochlich_Electrons_in_lattice_fields,1953_PR_LLP_Transformation,2010_alexandrov_advances_in_polaron_physics,2020_Devreese_Polaron_Lectures,2024_Arxiv_Demler_Review_Polaron,2024_PRA_Bipolaron_Perspective,2025_Arxiv_Massignan_Polaron_Review}. The interaction between the impurity and the surrounding bath particles dresses the impurity and causes a quasiparticle - characterized by an effective mass differing from that of the bare impurity and a finite residue or quasiparticle weight - to form. 
Studies \cite{2013_PRA_Schmidt_Field_Bose_Polaron,2015_PRL_Bruun_BEC_Impurity_Perturbation_Second_Order,2019_PRA_Bruun_Ladder,2015_PRA_Ardila_QMC_Perturbation} of neutral atom impurities 
suggest that the impurity-bath two-body scattering length $a_{\mathrm{IB}}$ is the key parameter -- irrespective of the condensate density $n_0$ -- determining the quasiparticle properties of the impurity. For $a_{\mathrm{IB}}<0$ ($a_{\mathrm{IB}}>0$), a stable (metastable) attractive (repulsive) polaron branch with a finite quasiparticle weight emerges.
The ability to modify the scattering length in ultracold atoms via Feshbach resonances has enabled the exploration of both polaron branches from weak to strong coupling regimes \cite{2016_PRL_Cornell_Bose_Polaron_Experiment,2016_PRL_Bruun_Bose_Polaron_Experiment,2020_Science_Zwierlien_Bose_Polaron}. 
In the vicinity of a scattering resonance, $a_{\mathrm{IB}}^{-1} \simeq 0$, the quasiparticle nature of the impurities becomes ill-defined \cite{2020_PRL_Perish_Repulsive_Fermi_Polaron,2022_Perish_Review_Repulsive_Bose_Polaron,2022_PRR_Pohl_Life_Death_of_Bose_Polaron,2024_Arxiv_Demler_Review_Polaron}, as the quasiparticle residue diminishes significantly.

The problem becomes more involved when the impurity-bath interaction becomes long-ranged, as for ionic \cite{2011_JLTP_Tempere_Ion_effective_mass_Frochlich,2018_PRL_IOnic_Impurity,2021_PRL_Bruun_Ionic_Polaron,2021_NC_astrakharchik_ionic_polaron,2024_PRA_Tempere_Path_Integral_Ionic_Polaron,2024_PRA_Krzysztof_Ionic_Polaron_BEC} or Rydberg atom \cite{2016_PRL_Schmidty_Mesoscopic_Rydberg,2018_PRA_Schmidt_Rydberg_Polaron_Theory,2018_PRL_Schmidt_Rydberg_polaron_experiment,2019_PRA_Shiva_Rydberg_Impurity,2022_Atoms_Shiwa_Rydberg} impurities. An ion interacts with surrounding bath particles via the long-range tail of the charge-neutral interaction \cite{2010_Nature_Ionic_Impurity,2005_Pethick_Static_Ionic_Impurity,2018_PRL_Meinert_Ionic_Impurity_BEC,2002_PRL_Lukin_Mesoscopic_ion}, whereas the interaction of a Rydberg impurity with a bath particle is proportional to the electron density and hence oscillatory and extending over large distances \cite{2000_PRL_Greene_Rydberg,2019_JPB_Meiles_Rydberg_Molecules}. Long-ranged interactions introduce an additional length scale, the interaction range $r_{\mathrm{eff}}$ of the impurity-bath potential, which makes the condensate density $n_0$ a significant factor in the quasiparticle characterization of the impurity \cite{2024_PRR_Aileen_Rydberg_phenomenology}.  
The filling parameter $r_\mathrm{eff}n_0^{1/3}$ quantifies the number of particles within the effective interaction volume. 
These particles can bind to the impurity, leading to a reshuffling of spectral weight from the polaron into multiply-occupied molecular states and challenging the quasiparticle description.  
These particles also effectively shield the interaction potential, modifying the interaction with distant bath particles and hence the quasiparticle dressing. 
As a result, a third length scale, the condensate coherence length $\xi \propto (a_{\mathrm{BB}} n_0)^{-1/2}$, where $a_{\mathrm{BB}}$ is the intraspecies scattering length, becomes relevant.

In this paper, we quantify the quasiparticle nature of long-ranged impurities as a function of the relevant length scales: the interparticle distance $n_0^{-1/3}$, the interaction range $r_{\mathrm{eff}}$, and the coherence length $\xi$. We calculate the quasiparticle weight $Z$ and effective mass $m_{\mathrm{eff}}$ across a range of interaction strengths and bath densities, showing how the interplay between these competing length scales leads to a range of different behavior. 
We employ two complementary theories in our investigation: a variational method using the coherent-state (CS) ansatz \cite{2014_PRA_Demler_Frochlich_Bose_polaron_Coherent_State,2016_PRL_Demler_Coherent_State_Beyond_Frohlich,2017_NJP_Grusdt_RG_MF_1D_Bose_Polaron,2019_PRA_Ens_Coherent_State_Ansatz} and the perturbative Gross-Pitaevskii (GP) theory \cite{2004_PRA_Pitaevskii_Moving_impurity_GP,2017_PB_Perturbative_GP}. 
We include the beyond Fr\"ohlich interaction terms to correctly treat both the strongly interacting regime and the existence of molecular bound states. Using the CS approach, we determine the saddle-point solution of the variational CS amplitudes in the extended Bogoliubov-Fr\"ohlich Hamiltonian, which allows for calculating all relevant quasiparticle properties. 
In the perturbative GP theory for a moving impurity, we closely follow the methodology outlined in \cite{2004_PRA_Pitaevskii_Moving_impurity_GP}, but improve the GP energy functional by incorporating the impurity-condensate perturbation interaction term, enabling us to explore the beyond Fr\"ochlich regime. Moreover, we compare these approaches with mean-field (MF) solutions and establish the range of validity for each method. 

Our main findings are as follows. When the coupling parameter $|a_{\mathrm{IB}}|n_0^{1/3}$ is of order unity or smaller,  ($|a_{\mathrm{IB}}|n_0^{1/3}\lesssim 1$), the impurity has a quasiparticle nature -- it possesses a finite quasiparticle weight and well-defined effective mass. 
This description fails at large coupling parameter $|a_{\mathrm{IB}}|n_0^{1/3}\gg 1$, i.e. near a scattering resonance.  
However, as the filling parameter $r_\mathrm{eff}n_0^{1/3}$ increases, the quasiparticle weight of a long-range impurity is progressively suppressed, and becomes arbitrarily small when $r_\mathrm{eff}n_0^{1/3}\gg 1$. This happens due to the increasing population of many molecular states. 

We also find that, in general, stronger intraspecies interactions can modify polaron properties by enhancing the quasiparticle weight and reducing the effective mass. At high condensate densities or strong intraspecies interactions, when $\xi \lesssim r_{\mathrm{eff}}$ is satisfied, the presence of the repulsive Bose field effectively modifies the impurity-bath potential. Thus, forming a bound state requires a stronger attraction between the impurity and bath particles, which leads to a shift of the scattering resonances towards deeper regions of the bare impurity-bath potential. By comparing three different impurity types, we demonstrate that these findings are generic: the impurity's nature influences only the two-body details, not the qualitative features of the polaron or its many-body dressing.

This paper is organized as follows. Section \ref{sec: Methodology} explains the general theoretical framework, the CS ansatz, and the perturbative GP method. Zero-range, ionic, and Rydberg impurities are discussed in sections \ref{sec: zero-range}, \ref{sec: Ionic RESULTS AND DISCUSSION}, and \ref{sec: Rydberg RESULTS AND DISCUSSION}, respectively. Section \ref{sec: Effect of Condensate Particle Interactions} presents the effect of bath-bath interactions. Finally, we conclude in Section \ref{sec: Conclusion}.

\section{\label{sec: Methodology} THEORY}
In this section, we describe the generic impurity-bath system before outlining the two methods used to compute this system's properties.  
\subsection{Impurity-Bath System}
We consider a single impurity of mass $m_{\mathrm{I}}$ inside of a weakly interacting Bose condensate composed of atoms of mass $m_{\mathrm{B}}$.
The impurity moves with momentum $\mathbf{p}_{\mathrm{I}} = m_{\mathrm{I}} \mathbf{V}$ and interacts with the bath particles through a central potential $U_{\mathrm{IB}}(\mathbf{r})$ (in real space) or $\mathcal{U}_{\mathrm{IB}}(\mathbf{k})$ (in momentum space).
This potential can be characterized by its zero-energy scattering length $a_{\mathrm{IB}}$ and interaction range $r_\mathrm{eff}$. While the former has a precise definition, the interaction range is not so well-defined and is taken to be roughly proportional to the physical extent of the interaction. The interaction range is quantitatively specified at the introduction of the relevant sections.

The interactions between the bath particles are considered to be contact potentials with coupling constant $g_{\mathrm{BB}} = 4\pi \hbar^2 a_{\mathrm{BB}}/m_{\mathrm{B}}$, where $a_{\mathrm{BB}}$ is the scattering length. The coherence length of the condensate is $\xi = \frac{\hbar}{\sqrt{2m_{\mathrm{B}}g_{\mathrm{BB}}n_0}}$.  
Another important quantity is the gas parameter, $\zeta=\sqrt{n_{\mathrm{loc}} a_{\mathrm{BB}}^3}$, where $n_{\mathrm{loc}} \approx \Delta N/r_\mathrm{eff}^3$ and $\Delta N$ is the number of particles localized within the interaction range of the impurity. Our theoretical methods assume that only a fractional proportion of phonons are excited from the uniform condensate, $\zeta\lesssim 0.1$. 

In momentum space, the most general form of the Hamiltonian describing the bosonic bath with a single impurity is
\begin{eqnarray}
    \hat{\mathcal{H}} &=&  \sum_{\mathbf{k}} \epsilon_{\mathbf{k}} \hat{a}_{\mathbf{k}}^{\dagger} \hat{a}_{\mathbf{k}} + \frac{g_{\mathrm{BB}}}{2\mathcal{V}} \sum_{\mathbf{k},\mathbf{k'},\mathbf{q}} \hat{a}_{\mathbf{k}+\mathbf{q}}^{\dagger} \hat{a}_{\mathbf{k'}-\mathbf{q}}^{\dagger} \hat{a}_{\mathbf{k'}} \hat{a}_{\mathbf{k}} \nonumber \\
    &+&\frac{\mathbf{\hat{p}_{\mathrm{I}}}^2}{2m_{\mathrm{I}}}+ \frac{1}{\mathcal{V}}\sum_{\mathbf{k},\mathbf{k'},\mathbf{q}} \mathcal{U}_{\mathrm{IB}}(\mathbf{q}) e^{-i \mathbf{q} \cdot \mathbf{\hat{R}}_{\mathrm{I}}}\hat{a}_{\mathbf{k}+\mathbf{q}}^{\dagger} 
    \hat{a}_{\mathbf{k}} , 
\label{General_Total_Hamiltonian}    
\end{eqnarray}
where $\hat{a}_{\mathbf{k}}$ 
is the annihilation operator for the bath 
particles
. Here $\epsilon_{\mathbf{k}} = \hbar^2 k^2/2m_{\mathrm{B}}$ is the free particle dispersion, $\mathcal{V}$ is the system volume, $\hat{\mathbf{R}}_{\mathrm{I}}$ and $\mathbf{\hat{p}_{\mathrm{I}}}$ are the impurity position and momentum operators, respectively. 
The first line of Eq. \eqref{General_Total_Hamiltonian} includes the kinetic and intraspecies interaction energy of the bath particles, whereas the second line contains the kinetic energy of the single impurity and its interactions with the bath particles. 
In the absence of the impurity, the condensate is assumed to be homogeneous with a density $n_0 = N/\mathcal{V}$, where $N$ is the total number of bath particles. 
All of our results are valid for any mass ratio $m^* \equiv m_{\mathrm{I}}/m_{\mathrm{B}}$, but we choose $m^* = 1$ for any numerical illustrations.

\subsection{\label{subsec:coherent state method} Coherent State Ansatz in Momentum Space}
\subsubsection{Extended Bogoliubov-Fröhlich Hamiltonian}
The extension of the CS ansatz formalism to long-range impurities, and the calculation of their effective mass require a careful treatment of the equations determining the coherent state amplitudes. 
Therefore, we briefly revisit the derivation of the extended Bogoliubov-Fr\"ohlich Hamiltonian from Eq.\,\eqref{General_Total_Hamiltonian} \cite{2016_PRL_Demler_Coherent_State_Beyond_Frohlich,2021_PRL_Bruun_Ionic_Polaron,2023_Scipost_Santos_Dipolar_Polaron}. 
Assuming a weak interaction among the bath particles, $n_0 a_{\mathrm{BB}}^3 \ll 1$, only a fraction occupies the non-zero momentum states. 
Therefore, we include interaction terms only up to second order in $\hat{a}_{\mathbf{k}}$ by applying the Bogoliubov transformation, where $\hat{b}_{\mathbf{k}}= u_{\mathbf{k}} \hat{a}_{\mathbf{k}}+v_{\mathbf{k}} \hat{a}_{\mathbf{-k}}^{\dagger}$ are Bogoliubov operators with $u_{\mathbf{k}},v_{\mathbf{k}} = \sqrt{\frac{1}{2} \left( \frac{\epsilon_k + g_{\mathrm{BB}}n_0}{E_k} \pm 1\right)}$ and the Bogoliubov dispersion is $E_k = \sqrt{\epsilon_k(\epsilon_k+2g_{\mathrm{BB}}n_0)}$. We additionally move into the frame co-moving with the impurity by applying the Lee-Low-Pines (LLP) transformation $\hat{\mathcal{H}}_{\mathrm{LLP}} = \hat{\mathcal{S}}^{-1} \hat{\mathcal{H}} \hat{\mathcal{S}}$, where $\hat{\mathcal{S}} = \exp{-i \hat{\mathbf{R}}_{\mathrm{I}} \cdot \hat{\mathbf{p}}_{\mathrm{B}}/\hbar}$, $\hat{\mathbf{p}}_{\mathrm{B}} = \sum_{\mathbf{k}} \hbar \mathbf{k} \hat{b}_{\mathbf{k}}^{\dagger} \hat{b}_{\mathbf{k}}$ is the total phonon momentum. Together, these transformations yield the LLP Hamiltonian $\hat{\mathcal{H}}_{\mathrm{LLP}} $ written in terms of Bogoliubov operators:

\begin{eqnarray} \label{LLP_Hamiltonian}   
    &\hat{\mathcal{H}}_{\mathrm{LLP}}& =  \frac{(\mathbf{{p}}_{\mathrm{tot}}-\mathbf{\hat{p}}_{\mathrm{B}})^2}{2m_{\mathrm{I}}}+\frac{g_{\mathrm{BB}}N^2}{2V} + n_0 \mathcal{U}_{\mathrm{IB}}(0) \nonumber \\
    &+& \sum_{\mathbf{k}} E_{\mathbf{k}} \hat{b}_{\mathbf{k}}^{\dagger} \hat{b}_{\mathbf{k}} 
    + \frac{\sqrt{N}}{\mathcal{V}} \sum_{\mathbf{k\ne 0}}  \mathcal{U}_{\mathrm{IB}}(\mathbf{k}) W_{\mathbf{k}} (\hat{b}_{\mathbf{k}}^{\dagger} + \hat{b}_{\mathbf{-k}})  \nonumber \\
    &+& \frac{1}{\mathcal{V}}\sum_{\mathbf{k},\mathbf{k'}\ne 0} \Bigg[ \mathcal{U}_{\mathrm{IB}}(\mathbf{k}-\mathbf{k'}) V_{\mathbf{k}\mathbf{k'}}^{(1)} \hat{b}_{\mathbf{k}}^{\dagger} \hat{b}_{\mathbf{k'}}  \\
    &\ & \ \ \ \ \ \ \ \ \ +  \mathcal{U}_{\mathrm{IB}}(\mathbf{k}+\mathbf{k'}) \frac{V_{\mathbf{k}\mathbf{k'}}^{(2)}}{2} \left( \hat{b}_{\mathbf{k'}}^{\dagger} \hat{b}_{\mathbf{k}}^{\dagger} + \hat{b}_{\mathbf{-k}}\hat{b}_{\mathbf{-k'}}\right) \Bigg] \nonumber.  
\end{eqnarray}
The first two lines of Eq.\,\eqref{LLP_Hamiltonian} yield the Fr\"ohlich Hamiltonian, whereas the latter terms constitute the beyond-Fr\"ohlich interactions. The total system momentum  $\mathbf{p_{\mathrm{tot}}}$ is a conserved quantity, and $W_{\mathbf{k}}= \sqrt{\epsilon_k/E_k}$, and $V_{\mathbf{k}\mathbf{k'}}^{(1,2)}= (W_k W_{k'} \pm W_k^{-1} W_{k'}^{-1})/2$ are the coupling coefficients partially determining the phonon-impurity interactions at Fr\"ohlich and beyond-Fr\"ohlich levels, respectively.
\subsubsection{Time-dependent variational coherent states}
In the ground state of the Fr\"ohlich Hamiltonian, $\mathbf{p_{\mathrm{tot}}}=0$, and the system consists of independent phonon modes $\hat{b}_{\mathbf{k}}$. 
Each mode can be solved by its coherent state, and they eventually yield a many-body eigenstate in the form $\ket{\Psi_{\mathrm{coh}}}  =  e^{-i \phi} e^{\sum_{\mathbf{k}} \beta_{\mathbf{k}}\hat{b}_{\mathbf{k}}^{\dagger} - \beta_{\mathbf{k}}^{*}\hat{b}_{\mathbf{k}}}$, where the coherent state amplitudes are $\beta_{\mathbf{k}} = -\frac{\sqrt{N} \mathcal{U}_{\mathrm{IB}}(\mathbf{k}) W_{\mathbf{k}}}{\mathcal{V} E_{\mathbf{k}}}$. 
This exact solution is employed to construct the CS ansatz for the beyond-Fr\"ohlich model. 
In this formulation, the coherent state amplitudes $\beta_{\mathbf{k}}(t) \in \mathbb{C}$ are treated as time-dependent variational parameters. The variational analysis over $\beta_{\mathbf{k}}^{*}(t)$, i.e. $i \hbar \frac{\partial \beta_{\mathbf{k}}}{\partial t} = \frac{\delta \bra{\Psi_{\mathrm{coh}}} \hat{\mathcal{H}}_{\mathrm{LLP}} \ket{\Psi_{\mathrm{coh}}}}{\delta \beta_{\mathbf{k}}^{*}}$, yields the equations of motion for the variational coherent state amplitudes:

\begin{eqnarray}
    i \hbar \frac{\partial \betak }{\partial t} &=& \left( E_k + \frac{\hbar^2 k^2 }{2 m_{\mathrm{I}}}-\frac{\hbar \mathbf{k} \cdot (\mathbf{p}_{\mathrm{tot}}- \sum_{\mathbf{k}} \hbar \mathbf{k} |\betak{}|^2)}{m_{\mathrm{I}}}\right) \betak \nonumber \\
    &+& \frac{\sqrt{N}}{\mathcal{V}} \UIBk{k} W_{\mathbf{k}} \nonumber\\
    &+& \frac{1}{\mathcal{V}} \sumk{k'} \bigg( \UIBk{k-k'} V_{\mathbf{k k'}}^{(1)} \betakk{k'} \nonumber \\ 
    & &\ \ \ \ \ \ \ \ \ \ \ + \UIBk{k+k'} V_{\mathbf{k k'}}^{(2)} \betakk{k'}^{*} \bigg).
    \label{time_dependent_betak_eqns}
\end{eqnarray}

\subsubsection{Saddle point solution}
The saddle point solution is found by setting $\partial_t \betak=0$, and the $\betak$ parameters satisfy the following equation
\begin{eqnarray}    \label{time_independent_betak_eqns}
    0 &=& \Omega_{\mathbf{k}} \betak + \frac{\sqrt{N}}{\mathcal{V}} \UIBk{k} W_{\mathbf{k}} \\
    &+& \frac{1}{\mathcal{V}} \sumk{k'}  \UIBk{k-k'} V_{\mathbf{k k'}}^{(1)}\betakk{k'} + \UIBk{k+k'} V_{\mathbf{k k'}}^{(2)}  \betakk{k'}^*, \nonumber
\end{eqnarray}
where $\Omega_{\mathbf{k}}[\mathbf{p}_{\mathrm{B}}] \equiv E_k + \frac{\hbar^2 k^2 }{2 m_{\mathrm{I}}}-\frac{\hbar \mathbf{k} \cdot (\mathbf{p_{\mathrm{tot}}}-\mathbf{p}_{\mathrm{B}})}{m_{\mathrm{I}}}$ and $\mathbf{p}_{\mathrm{B}} \equiv \sumk{k} \hbar \mathbf{k} |\betak|^2$. Note that $\UIBk{k}$ is spherically symmetric for central impurity-bath interactions, and the coupling coefficients $W_{\mathbf{k}}$ and $V_{\mathbf{kk'}}^{(1,2)}$ depend only on the magnitude $|\mathbf{k}|$ or $|\mathbf{k'}|$. 
For the ground-state properties of the polaron, such as its energy and residue, the total system momentum $\ptot$ is zero. 
This ensures that the spherically symmetric coherent state amplitudes satisfy $\betak \in \mathbb{R}$ and $\betakk{k}= \betakk{|k|}$. 
These symmetries simplify Eq.\eqref{time_independent_betak_eqns} to
\begin{eqnarray}\label{symmetric_time_independent_betak_eqns}
    0 &=& \Omega_{\mathbf{k0}} \betak + \frac{\sqrt{N}}{\mathcal{V}} \UIBk{k} W_{\mathbf{k}} \nonumber \\ 
    &+& \frac{W_{\mathbf{k}}}{\mathcal{V}} \sumk{k'}  \UIBk{k-k'}  W_{\mathbf{k'}}  \betakk{k'}, 
\end{eqnarray}
where $\Omega_{\mathbf{k0}} \equiv E_k + \frac{\hbar^2 k^2 }{2 m_{\mathrm{I}}}$. When the impurity-bath interaction is zero-ranged, $\UIBk{k} = g_{\mathrm{IB}}$ and Eq.\,\eqref{symmetric_time_independent_betak_eqns} possesses an analytical solution \cite{2016_PRL_Demler_Coherent_State_Beyond_Frohlich}.
For long-range impurities, the presence of $\UIBk{k\pm k'}$ terms in Eqs.~\eqref{time_independent_betak_eqns} and \eqref{symmetric_time_independent_betak_eqns} makes an analytical treatment intractable, and therefore we solve the set of linear equations for $\betak$ numerically, as described in the Appendix. \ref{Appendix:Numerics}.

\subsubsection{Calculation of observables}
The polaron energy can be obtained by calculating $\bra{\Psi_{\mathrm{coh}}} \hat{\mathcal{H}}_{\mathrm{LLP}} \ket{\Psi_{\mathrm{coh}}}$ at the saddle point \eqref{time_independent_betak_eqns}. The polaron energy and the residue for a finite total system momentum become \cite{2016_PRL_Demler_Coherent_State_Beyond_Frohlich,2023_Scipost_Santos_Dipolar_Polaron} 
\begin{eqnarray}\label{Polaron_energy_equations}
    E_{\mathrm{pol}} &=& n_0 \UIBk{0} + \frac{\sqrt{N}}{\mathcal{V}} \sumk{k} \UIBk{k} W_{\mathbf{k}} \mathrm{Re}[\betak] \nonumber \\
    &+& \frac{(\mathbf{p}_{\mathrm{tot}}-\mathbf{p}_{\mathrm{B}})^2}{2m_{\mathrm{I}}}+\sumk{k} \hbar \mathbf{k} \cdot \frac{(\mathbf{p}_{\mathrm{tot}}-\mathbf{p}_{\mathrm{B}})}{m_{\mathrm{I}}} |\betak|^2, \\
    Z &=& e^{-\frac{1}{2}\sumk{k}|\betak|^2} = e^{ -\frac{\mathcal{V}}{(2\pi)^2} \int dk \ k^2 |\beta_k|^2}, \label{residue_equations}
\end{eqnarray}
where the quasiparticle weight is given by $Z = |\braket{0}{\Psi_{\mathrm{coh}}(t \rightarrow \infty)}|$. The first term in Eq.\,\eqref{Polaron_energy_equations} is the mean-field (MF) energy $E_{\mathrm{MF}} = n_0 \int d^3 \mathbf{r} U_{\mathrm{IB}}(\mathbf{r})$, where the impurity is regarded as a perturbation to the uniform Bose field $\sqrt{n_0}$. The summation over the $\mathbf{k}$ vectors is executed by the integration $\frac{1}{\mathcal{V}}\sum_{\mathbf{k}} \rightarrow \frac{1}{(2\pi)^3} \int d^3 \mathbf{k}$, i.e. $\mathcal{V} = (2\pi / dk)^3$.

We compute the polaron effective mass through a perturbative expansion of the polaron energy in Eq.\,\eqref{Polaron_energy_equations}, assuming a small impurity velocity $\mathbf{V} = (\mathbf{p_{\mathrm{tot}}-\mathbf{p}_{\mathrm{B}}})/m_{\mathrm{I}}$. To this end, we expand the solution for $\betak$ parameters around the zero velocity solution, denoted by $\betakk{k0}$, in powers of $\mathbf{V}$:  

\begin{eqnarray} \label{betak_low_velocity_expansion}
    \betak &=& \betakk{k0}\left(1+\alpha_k \gammakv + \gamma_k \gammakvSquare\right),  \\
    |\betak|^2 &=& |\betakk{k0}|^2\left(1+2\alpha_k \gammakv + (\alpha_k^2 + 2\gamma_k) \gammakvSquare\right).\nonumber
\end{eqnarray}
The equations satisfied by $\betakk{k0}, \alpha_k$, and $\gamma_k$ on the zeroth, first, and second orders of $\mathbf{V}$ are given in Appendix \ref{Appendix:Effective Mass Equations}. Once the zero momentum solution $\betakk{k0}$, and the order corrections $\alpha_k$ and $\gamma_k$ are calculated for the impurity type, the corrections to the polaron energy \eqref{Polaron_energy_equations} proportional to $V^2$ become
\begin{eqnarray} \label{CS_velocity_dependent_energy}
    \frac{1}{2} m_{\mathrm{eff}} V^2 &=& \frac{1}{2}m_{\mathrm{I}} V^2 + \sumk{k} 2\alpha_k \betakk{k0}^2 \gammakvSquare \nonumber \\ &+& \sqrt{\frac{n_0}{\mathcal{V}}} \sumk{k} W_k \mathcal{U}_{\mathrm{IB}}(\mathbf{k}) \betakk{k0} \gamma_k \gammakvSquare .
\end{eqnarray}
This approach allows the calculation of the effective mass of the impurity solely through the zero-momentum solution, and the order corrections, without finding the finite-velocity solution of Eq.\eqref{time_independent_betak_eqns} explicitly. The contributions to the induced mass $m_{\mathrm{ind}} = m_{\mathrm{eff}}-m_{\mathrm{I}} $ in Eq.\eqref{CS_velocity_dependent_energy} are
\begin{eqnarray}
    m_{\mathrm{ind,1}} &=& \frac{4 m_{\mathrm{I}}\mathcal{V}}{3 \pi^2} \int dk \ k^2 \alpha_k \frac{\epsilon_k}{m^*} \betakk{k0}^2,  \\
    m_{\mathrm{ind,2}} &=& \frac{2 m_{\mathrm{I}}\sqrt{n_0 \mathcal{V}}}{3 \pi^2} \int dk \ k^2 W_k \UIBk{k} \gamma_k \frac{\epsilon_k}{m^*} \betakk{k0}, \nonumber
\end{eqnarray}
where the velocity $\mathbf{V} = V \mathbf{e_z}$ is assumed during integrations without loss of generality. 
\subsection{\label{subsec:GP method} Perturbative Gross-Pitaevskii Theory}
\subsubsection{Gross-Pitaevskii functional}
A complementary method to study the same impurity-bath problem is the theoretical framework of \cite{2004_PRA_Pitaevskii_Moving_impurity_GP}, wherein the total Gross-Pitaevskii (GP) energy is perturbatively expanded around a uniform condensate field. Here we closely follow this approach, but explicitly incorporate the previously neglected impurity-condensate interaction term, enabling us to explore the beyond-Fr\"ohlich regime.

We assume a GP ansatz $\Psi(\mathbf{r},t)$ on the condensate field and treat the impurity as an external potential moving with velocity $\mathbf{V}$. The total GP energy functional $K_{\mathrm{GP}} = E_{\mathrm{GP}} - \mu N$ of the system is
\begin{eqnarray} \label{GP_Energy_Functional}
    K_{\mathrm{GP}}&=& \int d^3\mathbf{r} \Bigg\{ \frac{\hbar^2}{2m_{\mathrm{r}}} |\nabla \Psi(\mathbf{r},t)|^2    \\ 
    &+& \left[U_{\mathrm{IB}}(\mathbf{r}-\mathbf{V}t)-\mu \right]|\Psi(\mathbf{r},t)|^2+\frac{g_{\mathrm{BB}}}{2} |\Psi(\mathbf{r},t)|^4\Bigg\},\nonumber
\nonumber\end{eqnarray}
where the chemical potential $\mu = g_{\mathrm{BB}}n_0$ is set to ensure a uniform condensate field far away from the impurity, and $m_{\mathrm{r}}$ is the reduced mass. The corresponding time-dependent GP equation (GPE) is obtained by a variational analysis $i \hbar \frac{\partial \Psi}{\partial t} = \frac{\delta K_{\mathrm{GP}}}{\delta \Psi^*}$: 
\begin{eqnarray} \label{GP_Equation}
    i \hbar \frac{\partial \Psi(\mathbf{r},t)}{\partial t} &=& -\frac{\hbar^2}{2m_{\mathrm{r}}} \nabla^2 \Psi(\mathbf{r},t)+ g_{\mathrm{BB}}|\Psi(\mathbf{r},t)|^2 \Psi(\mathbf{r},t) \nonumber \\ &+& \left[U_{\mathrm{IB}}(\mathbf{r}-\mathbf{V}t)-\mu \right]\Psi(\mathbf{r},t).
\end{eqnarray}
This GPE differs slightly from the one in \cite{2004_PRA_Pitaevskii_Moving_impurity_GP}, where the bare bath particle mass $m_{\mathrm{B}}$ is used instead of $m_{\mathrm{r}}$ in \eqref{GP_Equation}, considering the heavy impurity limit $m_{\mathrm{I}} \rightarrow \infty$. Furthermore, the treatment of the impurity as a moving external potential $U_{\mathrm{IB}}(\mathbf{r}-\mathbf{V}t)$ differs from conventional GPE approaches, which typically treat the finite momentum of the impurity and total momentum of the system by explicitly incorporating the LLP terms $(\ptot - \mathbf{p}_B)^2/2m_{\mathrm{I}}$, where $\mathbf{p}_B= i \hbar \int d^3 \mathbf{r} \Psi^*(\mathbf{r}) \nabla \Psi(\mathbf{r})$  \cite{1962_Annals_Gross_Motion_of_Impurity,2022_SciPost_Schmidt_GP_Bose_Polaron,2020_PRR_Fleisheur_Exact_1D_Bose_Polaron}.
By treating the impurity as an external potential with a fixed velocity vector, one implicitly neglects possible variations in either the impurity or bath momentum during the time evolution. This approximation is equivalent to a mean-field treatment of the bath-bath interactions $\propto \mathbf{p}_B \cdot \mathbf{p}_B$ induced after the LLP coordinate transformation. 
The equivalence between the GPE \eqref{GP_Equation} and the LLP Hamiltonian with the GP energy functional \cite{1962_Annals_Gross_Motion_of_Impurity,2022_SciPost_Schmidt_GP_Bose_Polaron,2020_PRR_Fleisheur_Exact_1D_Bose_Polaron} for the 
saddle point analysis is derived in more detail in the Appendix. \ref{Appendix:GP Equation Derivation}.

Now we assume that the condensate field $\Psi(\mathbf{r},t)$ consists of fluctuations $\delpsirt$ around the uniform density $\phi_0 = \sqrt{n_0}$, such that $\Psi(\mathbf{r},t) = \phi_0 + \delta \psi(\mathbf{r},t)$. With this assumption, our approach strictly deviates from \cite{2022_SciPost_Schmidt_GP_Bose_Polaron,2021_PRA_Massignan_GP_Bose_Polaron}, which utilizes the exact ground state MF solution of the GPE \eqref{GP_Equation}, while neglecting fluctuations. Expanding \eqref{GP_Equation} to first order in $\delpsirt$, one obtains the GPE for $\delpsirt$ as
\begin{eqnarray} \label{GP_Equation_delpsi}
    i \hbar \frac{\partial \delpsirt}{\partial t} &=& -\frac{\hbar^2}{2m_{\mathrm{r}}} \nabla^2 \delpsirt + U_{\mathrm{IB}}(\mathbf{r}-\mathbf{V}t)\phi_0  \nonumber \\
    &+&  U_{\mathrm{IB}}(\mathbf{r}-\mathbf{V}t)\delpsirt \nonumber \\ 
    &+&g_{\mathrm{BB}}\phi_0^2(\delpsirt + \delpsirtconj),
\end{eqnarray}
where the second line represents the interactions, neglected in \cite{2004_PRA_Pitaevskii_Moving_impurity_GP}, between the impurity and the bath perturbations. 
This term is analogous to the beyond Fr\"ohlich-Bogoliubov terms in the Hamiltonian $\mathcal{H}_{\mathrm{LLP}}$ \eqref{LLP_Hamiltonian}.  
Applying the same perturbative expansion of the field $\Psi(\mathbf{r},t)$ for the GP energy functional $E_{\mathrm{GP}}$ in \eqref{GP_Energy_Functional}, we find in the frame co-moving with the impurity $\mathbf{r} \to \mathbf{r} - \mathbf{V}t$:
\begin{widetext}
\begin{eqnarray} \label{GP_Energy_Functional_delpsi}
    E_{\mathrm{GP}} = \int d^3\mathbf{r} &&\Bigg\{ \frac{\hbar^2}{2m_{\mathrm{r}}} |\nabla \delpsirt|^2 + \UIBr \phi_0^2 + \frac{g_{\mathrm{BB}}\phi_0^4}{2} 
    + \UIBr \left[ \phi_0\left(\delpsirt + \delpsirtconj\right) + |\delpsirt|^2 \right] \nonumber \\
    &+& \frac{g_{\mathrm{BB}}}{2} \bigg[2\phi_0^3(\delpsirt + \delpsirtconj) 
     +\phi_0^2(4|\delpsirt|^2+ \delta \psi^2(\mathbf{r},t) + \delta \psi^*{^2}(\mathbf{r},t))  \bigg]\Bigg\}. 
\end{eqnarray}
\end{widetext}
Note that one can obtain the same GPE \eqref{GP_Equation_delpsi} also from the variation of \eqref{GP_Energy_Functional_delpsi} over $\delpsirtconj$. 

\subsubsection{Solution in momentum space}
Assuming the density perturbation closely follows the impurity position $\delta \psi \equiv \delta \psi (\mathbf{r} - \mathbf{V}t)$ imposes a condition
\begin{equation} \label{delpsir_identity}
    \frac{\partial \delpsirt}{\partial t} = -\mathbf{V} \cdot \nabla \delpsirt.
\end{equation}
We define the Fourier transforms $\delpsir = \frac{1}{\sqrt{\mathcal{V}}} \sumk{k} \delpsik e^{i\mathbf{k} \mathbf{r}}$ and $\delpsik = \frac{1}{\sqrt{\mathcal{V}}} \int d^3 \mathbf{r} \delpsir e^{-i\mathbf{k}\mathbf{r}}$. Then, we rewrite the GPE \eqref{GP_Equation_delpsi} in momentum space using the identity in \eqref{delpsir_identity}. Combining the equations for $\delpsik$ and $\delpsikk{-k}$, we obtain the following set of linear equations for $\delpsik$:
\begin{widetext}
\begin{equation} \label{psikr_equations}
    \frac{\gammakv+\epsilon_{k_r}}{\sqrt{\mathcal{V}}} \UIBk{k} \phi_0 = [\gammakvSquare-E_{k_r}^2] \delpsik 
    + \frac{1}{\mathcal{V}} \sumk{k'} \big[g_{\mathrm{BB}}n_0 \UIBk{k+k'}   -(\gammakv+\epsilon_{k_r} +g_{\mathrm{BB}}n_0)\UIBk{k-k'} \big] \delpsikk{k'},
\end{equation}
\end{widetext}
where $\epsilon_{k_r} = \frac{\hbar^2 k^2}{2m_{\mathrm{r}}}$ is the free particle dispersion for the reduced mass and $E_{k_r} = \sqrt{\epsilonkr(\epsilonkr+2g_{\mathrm{BB}}n_0)}$ is the energy of the Bogoliubov excitations with a reduced mass. We recover the solution $\delpsik = g_{\mathrm{IB}} \frac{\phi_0}{\sqrt{\mathcal{V}}} \frac{\gammakv + \epsilonkr}{\gammakvSquare-\Ekr^2}$ from Ref.\cite{2004_PRA_Pitaevskii_Moving_impurity_GP} by neglecting the beyond Fr\"ohlich term (the summation term in \eqref{psikr_equations}) and using a contact potential.
\subsubsection{Calculation of observables}
We calculate the polaron energy and residue for the ground state by setting $\mathbf{V}=0$. 
The effective mass is determined using an approach analogous to the CS ansatz by expanding the GP energy around the solution $E_{\mathrm{GP,0}}$ at $\mathbf{V}=0$ to second order in $\mathbf{V}$, i.e. $E_{\mathrm{GP}} \approx E_{\mathrm{GP,0}} + \frac{1}{2}m_{\mathrm{ind}}V^2$. To this end, the $\delpsikk{k}$ in \eqref{psikr_equations} is also expanded around the zero velocity solution $\delpsikk{k0}$, with corrections $\alpha_k$ and $\gamma_k$ at the first and second order of $\mathbf{V}$, respectively. 
Details of the numerical solution for $\delpsikk{k0}$ and the effective mass corrections are given in Appendix. \ref{Appendix:Numerics} and \ref{Appendix:Effective Mass Equations}.

The polaron energy $E_{\mathrm{GP,0}}$ and the correction due to the small impurity velocity $\frac{1}{2} m_{\mathrm{ind}} V^2$ are calculated by expressing the GP energy terms in \eqref{GP_Energy_Functional_delpsi} in momentum space and applying the low-velocity expansions. 
We obtain
\begin{eqnarray} \label{GP_Polaron_Energy}
    E_{\mathrm{GP,0}} &=& \UIBk{0}n_0 + \frac{\phi_0}{\sqrt{\mathcal{V}}} \sumk{k} \UIBk{k} \delpsikk{k0}\nonumber \\
    &+& g_{\mathrm{BB}}n_0 \sumk{k} |\delpsikk{k0}|^2
    \end{eqnarray}
    and
    \begin{eqnarray}
    \label{GP_Polaron_Energy_corr}
    \frac{1}{2}m_{\mathrm{ind}}V^2 &=&\sumk{k} \Bigg[ 2 \alpha_k |\delpsikk{k0}|^2+ \frac{\phi_0 \UIBk{k} \gamma_k}{\sqrt{\mathcal{V}} } \delpsikk{k0} \nonumber \\ 
    &+& \left( 2\gamma_k + \alpha_k\right) g_{\mathrm{BB}}n_0|\delpsikk{k0}|^2\Bigg]\gammakvSquare, 
\end{eqnarray}
where the polaron energy is similar to the CS ansatz expression in \eqref{Polaron_energy_equations}, differing only by the inclusion of the MF interactions between the perturbed field $\delpsir$ and the uniform condensate $\phi_0$ on the second line of Eq. \eqref{GP_Polaron_Energy}.
\footnote{Note that we have neglected this term in the polaron energy calculation, as it has a negligible contribution to the polaron energy when the perturbative expansion of the condensate field is valid, i.e. when $\frac{1}{\mathcal{V}}\sumk{k}\delpsikk{k} \ll n_0$, but it diverges unphysically near the scattering resonance. }

There exist three significant contributions to the induced mass $m_{\mathrm{ind}} = m_{\mathrm{ind,1}} +m_{\mathrm{ind,2}} +m_{\mathrm{ind,3}} $:
\begin{eqnarray} \label{induced_mass_equations}
    m_{\mathrm{ind,1}} &=& \frac{16m_{\mathrm{r}}}{3}\frac{\mathcal{V}}{4\pi^2} \int dk \ k^2 \alpha_k \epsilonkr|\delpsikk{k0}|^2 ,\nonumber\\
    m_{\mathrm{ind,2}} &=& \frac{8m_{\mathrm{r}}}{3} \frac{\phi_0 \sqrt{\mathcal{V}}}{4 \pi^2} \int dk \ k^2 \UIBk{k} \epsilonkr \gamma_k \delpsikk{k0} ,\\
    m_{\mathrm{ind,3}} &=& g_{\mathrm{BB}}n_0 \frac{\hbar^2 \mathcal{V}}{3 \pi^2} \int dk\  k^4 \left( 2\gamma_k + \alpha_k^2\right) |\delpsikk{ko}|^2 . \nonumber
\end{eqnarray}
The first mass term arises from the perturbed field's kinetic energy, the second from impurity-phonon interactions $\int d^3 \mathbf{r} \UIBr \phi_0 \delpsir$, and the third from the condensate-perturbed field interaction. Similar to the coherent state ansatz, the residue is $Z = |\braket{\Psi(t\rightarrow \infty)}{\phi_0}|$, which yields 
\begin{eqnarray}
    Z &=& e^{-\frac{1}{2}\sumk{k}|\delpsikk{k0}|^2} = e^{ -\frac{\mathcal{V}}{(2\pi)^2} \int dk \ k^2 |\delpsikk{k0}|^2}.
\end{eqnarray}


\subsection{Discussion}

The CS and GP theories detailed above expand the bath-bath and impurity-bath interactions around a uniform background in a similar fashion. 
The interaction components in the Bogoliubov-Fr\"ohlich Hamiltonian \eqref{LLP_Hamiltonian} and the perturbative GP energy functional \eqref{GP_Energy_Functional_delpsi} are the momentum and real space representations of the same Bogoliubov expansion, with a slight difference in the definition of uniform background densities $\sqrt{n_0}$ and $\phi_0$, respectively. The $n_0$ in the $k$-space Hamiltonian represents the uniform field created by all the particles, including both the $k=0$ state and the excited states, i.e. $N= N_0 + \sumk{k} a_k^{\dagger} a_k $. In the GP theory, the uniform background and the corresponding chemical potential are fixed by $\phi_0$. This difference leads to the only qualitatively distinct term in Eq. \eqref{GP_Polaron_Energy}. 

The theories are also similar in their approach to the effective mass calculation. Both apply an expansion around a zero-velocity solution, leading to analytically tractable expressions. In the perturbative GP framework, the impurity is modeled as an external potential moving at a fixed velocity, which corresponds to an implicit mean-field treatment of the Bose–Bose interactions induced by the LLP transformation. By contrast, the CS ansatz retains all terms of the LLP Hamiltonian. This distinction manifests only in the time evolution of the system. Nevertheless, in both approaches the impurity is associated with a fixed momentum vector during the saddle-point analysis, an assumption that remains valid as long as the momentum remains nearly constant in the steady state.

The minor differences between the two approaches give rise to slight deviations in the quasiparticle properties reported below. In particular, the GP theory predicts slightly larger values of both the quasiparticle weight and the effective mass for any impurity type compared to the CS ansatz. Furthermore, the onset of the new bound state is predicted to occur at a deeper shift in the impurity-bath potential within the CS framework.

A notable limitation of the Bogoliubov treatment of bath-bath interactions in the Bose polaron problem is evidenced by the divergence of the polaron energy when the impurity-bath interaction supports a zero-energy scattering resonance. This divergence can be partially attributed to the underestimated energy cost of repulsive interactions among bath particles accumulated around the impurity. Despite this deficiency, the extended model treated with the CS ansatz remains a robust and effective method to explore the strongly coupled regime at finite interaction strengths \cite{2016_PRL_Demler_Coherent_State_Beyond_Frohlich,2024_Arxiv_Demler_Review_Polaron}.

In the following sections, we apply these methods to three different impurity-bath interactions. Dimensionless variables are used throughout. Length and energy dimensions, and the interaction range, specific to each impurity system, are specified in the introduction to the relevant sections. 

\begin{figure}[t]
    \includegraphics[width=8cm,keepaspectratio]{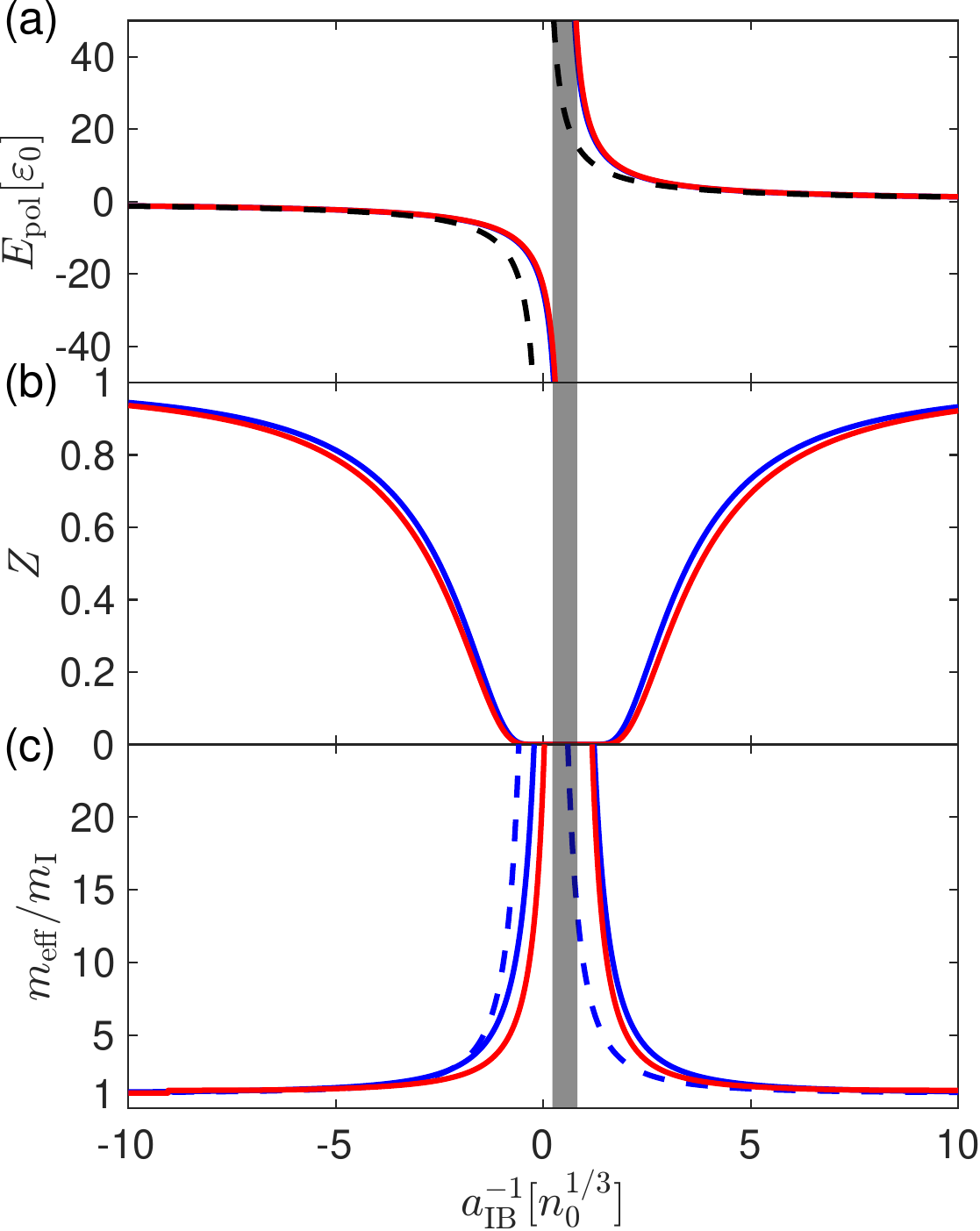}
    \caption{ \justifying The quasiparticle properties of the short-range impurity: \textbf{(a)} the polaron energy $E_{\mathrm{pol}}$, \textbf{(b)} the residue $Z$, and \textbf{(c)} the effective masses $m_{\mathrm{eff}}$, as a function of ${a}^{-1}_{\mathrm{IB}}$, calculated by the CS ansatz (red solid) and GP theory (blue solid). The black dashed line is the MF polaron energy $E_{\mathrm{MF}}$. The Fr\"ohlich effective mass (blue dashed) is also shown. The numerical computations are executed with the momentum cut-off $\Lambda n_0^{-1/3} = 30$, the step size $d{k} n_0^{-1/3}=0.1$, and $a_{\mathrm{BB}} = 0.01n_0^{-1/3}$. The gray shaded areas indicate the parameter regime in which the density around the impurity violates the small gas parameter assumption.\label{fig:Contact_Polaron_Properties_Figure}}
\end{figure}

\section{\label{sec: zero-range}Illustrative example: zero-range impurity}

We begin our analysis by considering a short-ranged impurity-bath interaction parameterized by a zero-range contact potential $U_{\mathrm{IB}}(\mathbf{r}) = g_{\mathrm{IB}} \delta (\mathbf{r})$. 
This system has already been studied extensively \cite{2024_Arxiv_Demler_Review_Polaron}, and thus here it serves as a benchmark for the employed theories, illustrates many of the qualitative features found in long-range impurities, and demonstrates how the saddle point solutions provide clear physical insights (See Appendix \ref{Appendix:Time-dependent and Saddle Point}).
It is also useful in that it is amenable to a fully analytical treatment, and we obtain known results for the polaron energy and quasiparticle weight as well as a new result for the effective mass incorporating beyond-Fr\"ohlich interactions.

To first order in the Born approximation, the coupling constant is $g_{\mathrm{IB}} = 2 \pi \hbar^2  a_{\mathrm{IB}}/m_{\mathrm{r}}$, where $a_{\mathrm{IB}}$ is the scattering length between the impurity and a bath particle. Whenever a renormalization is necessary, all the terms of the Dyson series are accounted for using \cite{pethick2008bose}:
\begin{equation} \label{gIB_Regulariation}
    \frac{1}{g_{\mathrm{IB}}} = \frac{m_{\mathrm{r}}}{2 \pi \hbar^2 a_{\mathrm{IB}}} - \frac{1}{\mathcal V}\sum_{\mathbf{k}} \frac{2m_{\mathrm{r}}}{\hbar^2 k^2}.
\end{equation}
The interparticle distance $n_0^{-1/3}$ is used as the length scale. The energy is scaled by $\varepsilon_0 \equiv \frac{\hbar^2 n_0^{2/3}}{2m_{\mathrm{r}}}$. 

The contact potential permits analytical solutions to be found for both the coherent state amplitudes $\betak$ and for perturbative fields $\delpsikk{k}$ for any $\ptot$ or $\mathbf{V}$ \footnote{The saddle-point solutions of $\betak$'s remain real-valued in the case of a short-range impurity, as long as the impurity moves within the subsonic regime \cite{2021_PRL_Demler_Bose_Polaron_Cherenkov}.}. 
By dividing Eq.~\eqref{time_independent_betak_eqns} through by $\Omega_{\mathbf{k}}$ and recognizing that the sum $\sumk{k} W_{\mathbf{k} }\betak$ is simply a scalar, one obtains $\betak$. Regularizing $g_{\mathrm{IB}}$ as in \eqref{gIB_Regulariation} gives the result first obtained in Ref.\,\cite{2016_PRL_Demler_Coherent_State_Beyond_Frohlich}
\begin{eqnarray} \label{betak_solutions_short_range}
    \betak = -\frac{\sqrt{N}}{\mathcal V} \frac{W_k}{\Omega_{\mathbf{k}}}  \frac{4\pi }{\aIBinverse-\aPlusinverse},
\end{eqnarray}
where $\aPlusinverse \equiv \frac{2\pi}{ \mathcal{V}} \sumk{k}\left( \frac{1}{k^2} -\frac{W_k^2}{\Omega_{\mathbf{k}}}\right)$. Then, one can use \eqref{betak_solutions_short_range} to calculate the ground state polaron energy \eqref{Polaron_energy_equations} and residue \eqref{residue_equations} analytically \cite{2016_PRL_Demler_Coherent_State_Beyond_Frohlich}: 
\begin{eqnarray} \label{contact_polaron_energy_CS}
    E_{\mathrm{pol}} &=& \frac{4\pi}{\aIBinverse-\aPlusinverse},\\
    Z &=& \exp\left[ -\frac{4}{3\sqrt{\pi a_{\mathrm{BB}}}} \frac{1}{\left( \aIBinverse-\aPlusinverse\right)^2}\right].
\end{eqnarray}
The order corrections $\alpha_k$ and $\gamma_k$ in the effective mass calculations also yield exact analytical expressions \eqref{short_range_order_corrections}, as shown in Appendix \ref{Appendix:Effective Mass Equations}.  This results in the effective mass:
\begin{eqnarray}  \label{CS_effective_mass_contact}
    \frac{m_{\mathrm{eff}}}{m_{\mathrm{I}}} = 1 + \frac{64}{45 \sqrt{\pi}\sqrt{a_{\mathrm{BB}}}} \frac{1}{(\aIBinverse-\aPlusinverse)^2},
\end{eqnarray}
where the contribution at the order of $(\aIBinverse-\aPlusinverse)^{-3}$ is neglected. For simplicity, we have set $m^* =1$, which yields $a_+^{-1} = \frac{8}{3\pi}\sqrt{16 \pi a_{\mathrm{BB}}} $. The general form of \eqref{CS_effective_mass_contact} and \eqref{GP_effective_mass_contact} for any mass ratio $m^*$ can be found in the Appendix. \ref{Appendix:Effective Mass Equations}. 

In the perturbative GP method, the analytical solution for the perturbed fields $\delpsikk{k}$ in \eqref{psikr_equations} is derived similarly to the coherent state method, yielding
\begin{eqnarray} \label{delpsik_solution_short_range}
    \delpsikk{k} = \frac{\phi_0}{\sqrt{\mathcal{V}}} \frac{\mathbf{k} \cdot \mathbf{V}+\epsilonkr}{(\mathbf{k} \cdot \mathbf{V})^2-\Ekr^2}\frac{1}{a_{\mathrm{IB}}^{-1} - \frac{1}{\mathcal{V}} \sumk{k'} \frac{ \mathbf{k'}\cdot \mathbf{V} + \epsilon_{k'_r}}{(\mathbf{k'}\cdot \mathbf{V})^2 - E_{k'_r}^2}}. \nonumber \\
\end{eqnarray}
 We evaluate \eqref{delpsik_solution_short_range} for $\mathbf{V}=0$ to calculate the polaron energy and residue, and use the induced mass expressions in \eqref{induced_mass_equations} to calculate the effective mass. The polaron energy and residue are
\begin{eqnarray}
    E_{\mathrm{GP}} &=& \frac{4\pi }{\aIBinverse - a_{\mathrm{+,GP}}^{-1}},\\
    Z_{GP} &=& \exp\left[ -\frac{\pi}{a_{\mathrm{+,GP}}^{-1}}\frac{1}{(\aIBinverse-a_{\mathrm{+,GP}}^{-1})^2}\right] ,
\end{eqnarray}
where $a_{\mathrm{+,GP}}^{-1} \equiv \sqrt{16\pi a_{\mathrm{BB}} m^*/(m^*+1)}$ is the scattering resonance shift calculated by the GP method. The total induced mass yields the following beyond Fr\"ohlich effective mass for the short-range impurity
\begin{eqnarray} \label{GP_effective_mass_contact}
    \frac{m_{\mathrm{eff}}}{m_{\mathrm{I}}} &=& 1+ \frac{2 \sqrt{2\pi}}{3\sqrt{a_{\mathrm{BB}}}}  \frac{5/8}{ (\aIBinverse-a_{\mathrm{+,GP}}^{-1})^2} ,
\end{eqnarray}
where, interestingly, $m_{\mathrm{ind,1}}$, $m_{\mathrm{ind,2}}$, and $m_{\mathrm{ind,3}}$ in \eqref{induced_mass_equations} all have the same functional form and differ only by the prefactors 1, -1/2, and 1/8, respectively. The result \eqref{GP_effective_mass_contact} is a significant improvement over the Fr\"ohlich counterpart  $m_{\mathrm{eff}}/m_{\mathrm{I}} = 1+ \frac{2\sqrt{\pi}}{3} \sqrt{a_{\mathrm{BB}}^3} \left(\frac{a_{\mathrm{IB}}}{a_{\mathrm{BB}}}\right)^2$, found in \cite{2004_PRA_Pitaevskii_Moving_impurity_GP}. This limit can be recovered from \eqref{GP_effective_mass_contact}, by ignoring the beyond Fr\"ohlich corrections $m_{\mathrm{ind,2}}$ and $m_{\mathrm{ind,3}}$, and taking the limit $a_{GP,+}^{-1} \rightarrow 0$. Interestingly, both approaches yield the same functional form for the effective mass, with the minor differences in the definition of scattering resonance shift $a_+$ and the prefactors leading to quantitative differences at the $5\%$ level. 

The quasiparticle properties computed by the CS ansatz (solid red line) and perturbative GP (solid blue line) theories are shown in Fig.\ref{fig:Contact_Polaron_Properties_Figure}. The MF energy of the polaron, $E_{\mathrm{MF}} = 4 \pi a_{\mathrm{IB}}$, is plotted as a dashed black curve. First, the polaron energy remains repulsive (attractive) on the positive (negative) scattering lengths, respectively, only shifted by $a_+$, resulting from the bath-bath particle interactions. The polaron energies increase in magnitude as the impurity-bath coupling strengthens. We note that both theories remain valid even in the regime of the strong coupling parameter. But they both diverge near the scattering resonance, which occurs partially due to the inadequate treatment of the bath-bath coupling around the impurity. This is also highlighted by the failure of both theories near the scattering resonances, indicated by the gray shaded regions in the plots. 

In Fig.\ref{fig:Contact_Polaron_Properties_Figure}(b), the suppression of the quasiparticle weight as the coupling parameter $a_{\mathrm{IB}}$ approaches the strong coupling regime is clearly demonstrated. But we note that the quasiparticle weight always recovers one as $a_{\mathrm{IB}} \to 0$. This behavior holds specifically for the zero-range impurity and manifests for the long-range impurity only in the dilute-density limit.
In Fig.\ref{fig:Contact_Polaron_Properties_Figure}(c), we show the effective mass results for both Fr\"ohlich and beyond-Fr\"ohlich models of both theories. The Fr\"ohlich and beyond-Fr\"ohlich effective masses exhibit symmetry around $a_{\mathrm{IB}}^{-1} = 0$ and $a_{\mathrm{IB}}^{-1}= a_+^{-1}$, respectively. In both cases, the effective mass converges to one as the impurity-bath coupling approaches zero. The results of the Fr\"ohlich model fully agrees with the results of the perturbation theory in \cite{2015_PRA_Ardila_QMC_Perturbation, 2024_Arxiv_Demler_Review_Polaron}. In the beyond Fr\"ohlich model, the shift of the resonance location towards a positive scattering length is qualitatively consistent with quantum Monte Carlo results \cite{2015_PRA_Ardila_QMC_Perturbation, 2024_Arxiv_Demler_Review_Polaron}. 

\section{\label{sec: results on long range impurites}Results: long-range impurities}
In this section, we present the results for long-range impurities near the intermediate filling density regime $r_{\mathrm{eff}}n_0^{1/3} \simeq 1$.
\subsection{\label{sec: Ionic RESULTS AND DISCUSSION} Ionic Impurity}

At long range, the interaction between an ion and a neutral atom is determined by the charge-neutral interaction $-\alpha/2r^4$, where $\alpha$ is the polarizability of one of the bath atoms. The long-range tail of the ionic impurity does not lead to a cutoff distance to define the interaction range, and thus we determine it via setting $\frac{\hbar^2}{2m_{\mathrm{r}} r_\mathrm{ion}^2 }= |U_{\mathrm{IB}}(r_\mathrm{ion})|$, which yields $r_\mathrm{ion}= \frac{\sqrt{ m_{\mathrm{r}} \alpha}}{\hbar}$.
At small $r$, the interaction changes due to the influence of the core electrons and becomes repulsive. This is typically modeled by regularizing the polarization potential, leading to (in real and momentum space)
\begin{eqnarray} \label{ionic_potential}
    U_{\mathrm{ion}}(\mathbf{r}) &=& - \frac{\alpha}{2(r^2+b^2)^2} \frac{r^2-c^2}{r^2+c^2}, \\
    \mathcal{U}_{\mathrm{ion}}(\mathbf{k}) &=&  -\frac{\alpha \pi^2}{2b} \frac{4bc^2(e^{-ck} - e^{-bk})-k(b^4-c^4) e^{-bk}}{(b^2-c^2)^2k}. \nonumber 
\end{eqnarray}
The parameter $b$ defines the depth of the ionic potential well, whereas the $c$ determines the position of the inner repulsive barrier. The length and energy scale of the problem are set as $r_{\mathrm{ion}}$ and $\varepsilon_{\mathrm{ion}} = \frac{\hbar^2}{2m_{\mathrm{r}} r_{\mathrm{ion}}^2}$. Here we fix $c = 0.0023$ to facilitate comparison with other literature results \cite{2021_PRL_Bruun_Ionic_Polaron}, and vary ${b} \in [0.16,1.05]$. In this range, the potential is deep enough to support a single bound state for $0.26 <{b}<0.58$ and two for $0.16<{b}<0.26$. 

Unlike the short-range impurity considered in the previous section, where the interaction range of the impurity-bath interaction is much smaller than the interparticle distance $n_0^{-1/3}$, the ratio of these length scales can be tuned as a function of density when the impurity-bath interaction is long-ranged.
In the low filling parameter limit defined by $r_{\mathrm{eff}} \ll n_0^{-1/3}$, the zero-energy scattering length $a_{\mathrm{IB}}$ remains a useful quantity, and the dimensionless parameter $n_0 a_{\mathrm{IB}}^3$ serves as a basis for guiding the analysis of the ionic polaron \cite{2024_PRR_Aileen_Rydberg_phenomenology}. 
We define the zero-range MF energy $E_{\mathrm{zr}} = 4\pi a_{\mathrm{IB}} n_0$, which the polaron energy is expected to approach in this regime. The MF energy is $E_{\mathrm{MF}} = \mathcal{U}_{\mathrm{ion}}(0)n_0$.

\begin{figure}[t]
    \includegraphics[width=8cm,keepaspectratio]{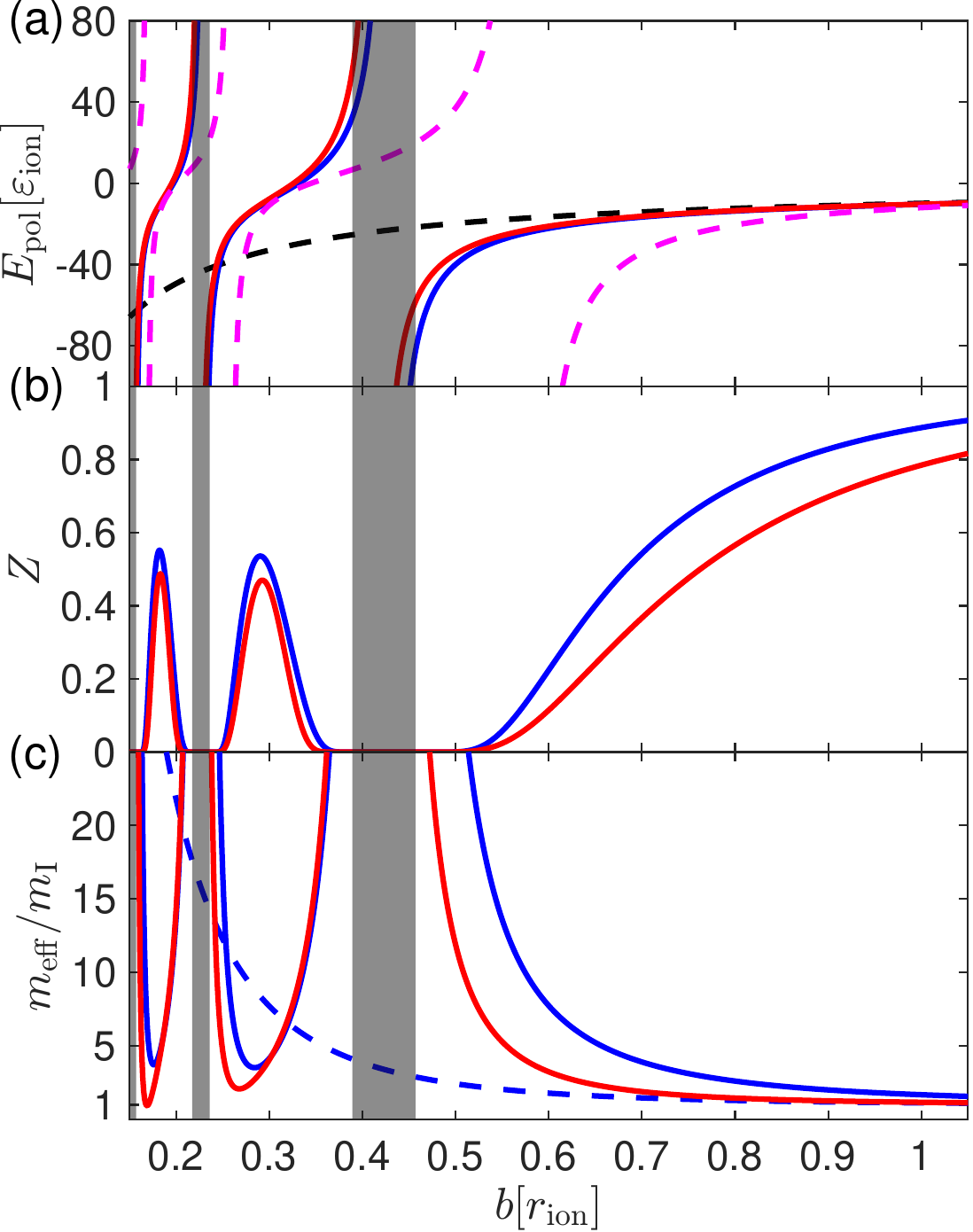}
    \caption{\justifying The quasiparticle properties of the ionic impurity calculated by the GP theory (blue solid) and CS ansatz (red solid): \textbf{(a)} The polaron energy $E_{\mathrm{pol}}$, \textbf{(b)} the quasiparticle weight $Z$, \textbf{(c)} the effective mass $m_{\mathrm{eff}}$, as a function of the well depth $b$ of the ionic potential. The dashed magenta (black) line in \textbf{(a)} represents $E_{\mathrm{zr}}$ ($E_{\mathrm{MF}}$). The blue dashed line is the Fr\"ohlich effective mass. We set $a_{\mathrm{BB}} = 0.05 r_{\mathrm{ion}}$ and $n_0 r_{\mathrm{ion}}^3 = 1$. The numerical parameters for the momentum are $dkr_{\mathrm{ion}}= 0.1$ and $\Lambda r_{\mathrm{ion}}= 30$. The gray shaded areas indicate where the small gas parameter assumption breaks down. \label{fig:Ionic_Polaron_Properties_Figure}}
\end{figure}

We illustrate the quasiparticle properties of the ionic impurity in Fig.\ref{fig:Ionic_Polaron_Properties_Figure}, computed by the GP (blue lines) and CS (red lines) theories, for $n_0 = 1$, along with the  $E_{\mathrm{zr}}$ (magenta dashed line) and $E_{\mathrm{MF}}$ (black dashed line) energy curves. The polaron energies calculated by the two methods, in mutual agreement, deviate from the mean-field energy $E_{\mathrm{MF}}$, which predicts a monotonic decrease in the attractive polaron energy as $U_{\mathrm{ion}}(\mathbf{r})$ deepens and does not predict the existence of the repulsive polaron branch. Rather, the polaron energy follows a qualitative trend similar to that of $E_{\mathrm{zr}}$. 

Before the interaction is strong enough to support a bound state, ${b}> 0.58$, an attractive polaron is formed, with the quasiparticle weight approaching unity as the ionic interaction weakens. Accordingly, the effective mass converges to $m_{\mathrm{I}}$. However, the Fr\"ohlich and beyond-Fr\"ohlich models start to differ drastically around ${b}\sim 0.5$, as the Fr\"ohlich treatment can not describe molecular formation. But both CS and GP results yield diverging polaron energy and effective mass expressions around the scattering resonance, due to the shortcomings of the Bogoliubov expansion over the uniform density.

A notable new branch emerges within the range ${b} \in [0.25,0.4]$, initially characterized by the repulsive polaron and subsequently transitioning to the attractive polaron. In this regime, $Z$ never approaches unity, as the spectral weight is partially lost to the molecular states. Yet the quasiparticle residue $Z$ obtains a well-defined finite value $\sim 0.58$ between the two scattering resonances, along with a well-defined effective mass. More precisely, the quasiparticle weight remains finite in the weak $|a_{\mathrm{IB}}|n_0^{1/3} \ll 1$ and intermediate $|a_{\mathrm{IB}}|n_0^{1/3} \simeq 1$ coupling parameter regimes, whereas it becomes vanishingly small for strong coupling parameter $|a_{\mathrm{IB}}|n_0^{1/3} \gg 1$.  Moreover, this behavior recurs with the introduction of the new bound states, such as at ${b} \sim 0.26$. Each recurrence, introduced by the scattering length $a_{\mathrm{IB}}$ varying from positive to negative infinity, echoes the transition from the repulsive to attractive polaron branch of the zero-range impurity. 

\subsection{\label{sec: Rydberg RESULTS AND DISCUSSION} Rydberg Impurity}

We now consider another type of long-range impurity, created by a Rydberg excitation of a ground-state atom inside a Bose condensate \cite{2016_PRL_Schmidty_Mesoscopic_Rydberg,2018_PRA_Schmidt_Rydberg_Polaron_Theory,2018_PRL_Schmidt_Rydberg_polaron_experiment,2024_PRR_Aileen_Rydberg_phenomenology}. 
The interaction between the Rydberg and the bath atoms is determined by the scattering of the Rydberg electron off of the bath atoms, and hence the electronic wavefunction and its spatial symmetries become key ingredients of the impurity-bath potential. Here we focus only on a spherically symmetric electronic $s-$state with the principal quantum number $n$, whose interaction potential can be written, in  real and momentum space \cite{2013_PRA_Rydberg_Fourier} respectively, as
\begin{eqnarray}
    &&U_{\mathrm{Ryd}}(\mathbf{r}) = 2 \pi {a}_{\mathrm{s}} |{\Psi}_{n00}(\mathbf{r})|^2, \\
    &&\mathcal{U}_{\mathrm{Ryd}}(\mathbf{k}) =  \pi {a}_{\mathrm{s}} \sum_{i, j=0}^{n-1} {n-1\choose {i}} {n-1\choose {j}} (-1)^{i+j}  \\
    &&\times{i + j +2\choose {i+1}} \frac{_2 F_1\left( -\frac{i +j+1}{2},-\frac{i+j}{2};\frac{3}{2};-(\frac{{k}n}{2})^2\right)}{\left[1+ (\frac{{k}n}{2})^2\right]^{i + j +2}}\nonumber.
\end{eqnarray}
The Bohr radius $a_0$ and the Hartree energy $\frac{\hbar^2}{m_ea_0^2}$ define the length and energy scale of the interaction. The electron mass is $m_e$, and $\Psi_{n00}(\mathbf{r})$ is the hydrogenic wave function. $ _2F_1(a,b;c;z)$ is the confluent hypergeometric function. 
The key parameters determining the strength and range of the potential are the electron-bath particle scattering length $a_{\mathrm{s}}$, and the typical extension of the Rydberg orbit  $r_{\mathrm{eff}} \simeq 2n^2$. Without loss of generality, we consider the bath atoms to be $^{87}$Rb and fix $n=50$,
We use the electron-atom scattering length $a_{\mathrm{s}}$ as the tuning parameter to investigate the quasiparticle properties. The first and subsequent bound states between the impurity and bath particle are supported when ${a}_{\mathbf{s}}$ decreases below $-0.05$, $-0.22$, $-0.49$, and so forth.  

\begin{figure}[t]
    \includegraphics[width=8cm,keepaspectratio]{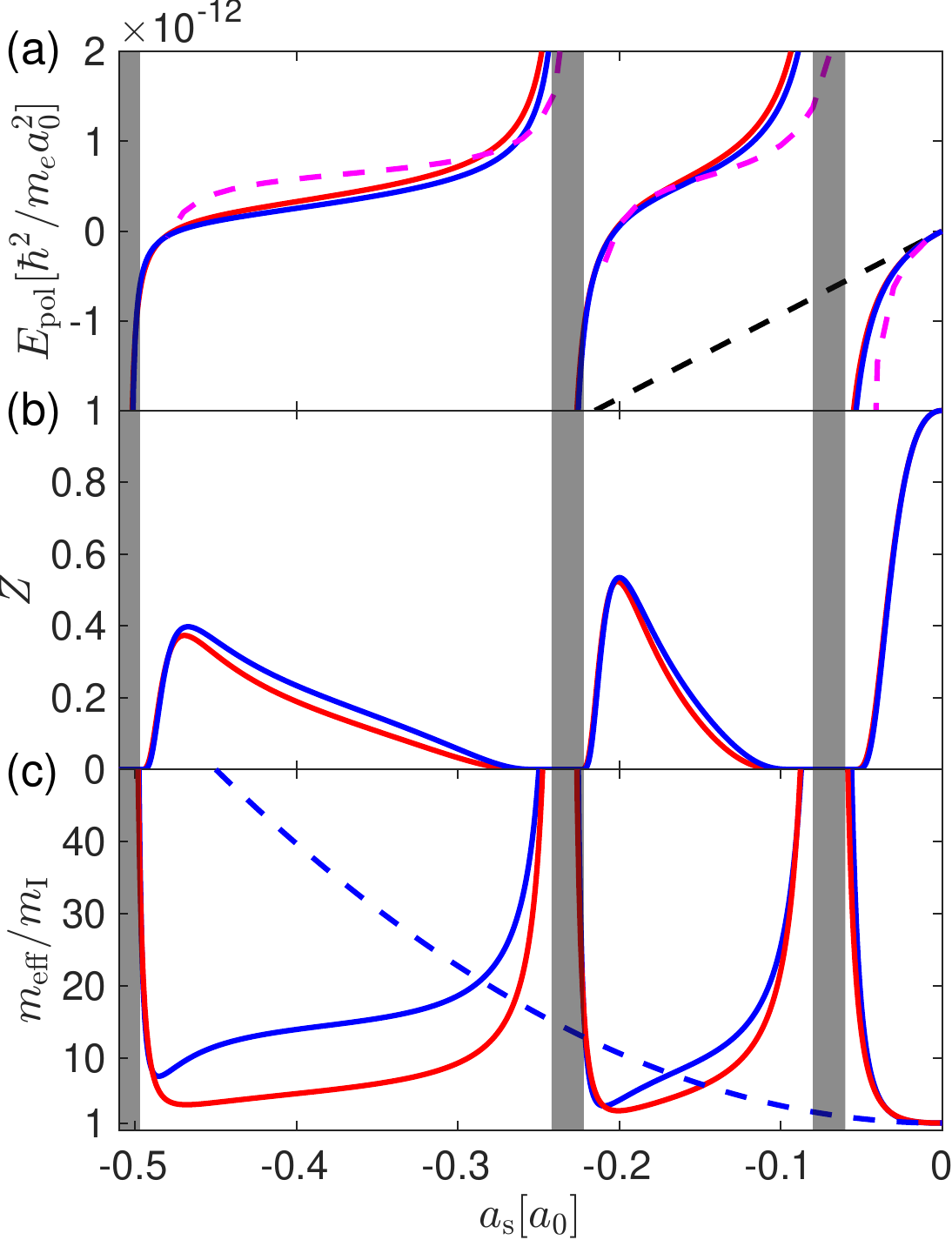}
    \caption{\justifying The quasiparticle properties of the Rydberg impurity in a weakly interacting Bose condensate, calculated by the CS ansatz (red lines) and the perturbative GP theory (blue lines): \textbf{(a)} The polaron energy $E_{\mathrm{pol}}$, \textbf{(b)} the quasiparticle weight $Z$, \textbf{(c)} the effective mass $m_{\mathrm{eff}}$, as a function of the electron-bath particle scattering length $a_{\mathrm{s}}$. The dashed magenta (black) line in \textbf{(a)} represents $E_{\mathrm{zr}}$ ($E_{\mathrm{MF}}$). The polaron energy, significantly deviating from $E_{\mathrm{MF}}$, is largely consistent with the $E_{\mathrm{zr}}$. The dashed blue line in \textbf{(c)} is the Fr\"ohlich effective mass.
    Here we set the bath-bath scattering length $a_{\mathrm{BB}} = 100a_0$ and the condensate density $n_0^{1/3} r_{\mathrm{eff}} \simeq 0.5$ ($n_0 a_0^3 = 1.48\times 10^{-12}$) 
    . In numerical analysis, we set $dka_0 = 10^{-5}$ and $\Lambda = 4000dk$. The gray shaded areas indicate where the small gas parameter assumption breaks down.  \label{fig:Rydberg_Polaron_Properties_Figure}}
\end{figure}

\begin{figure*}
    \begin{subfigure}{0.445\textwidth}
        \includegraphics[width = \textwidth,keepaspectratio]{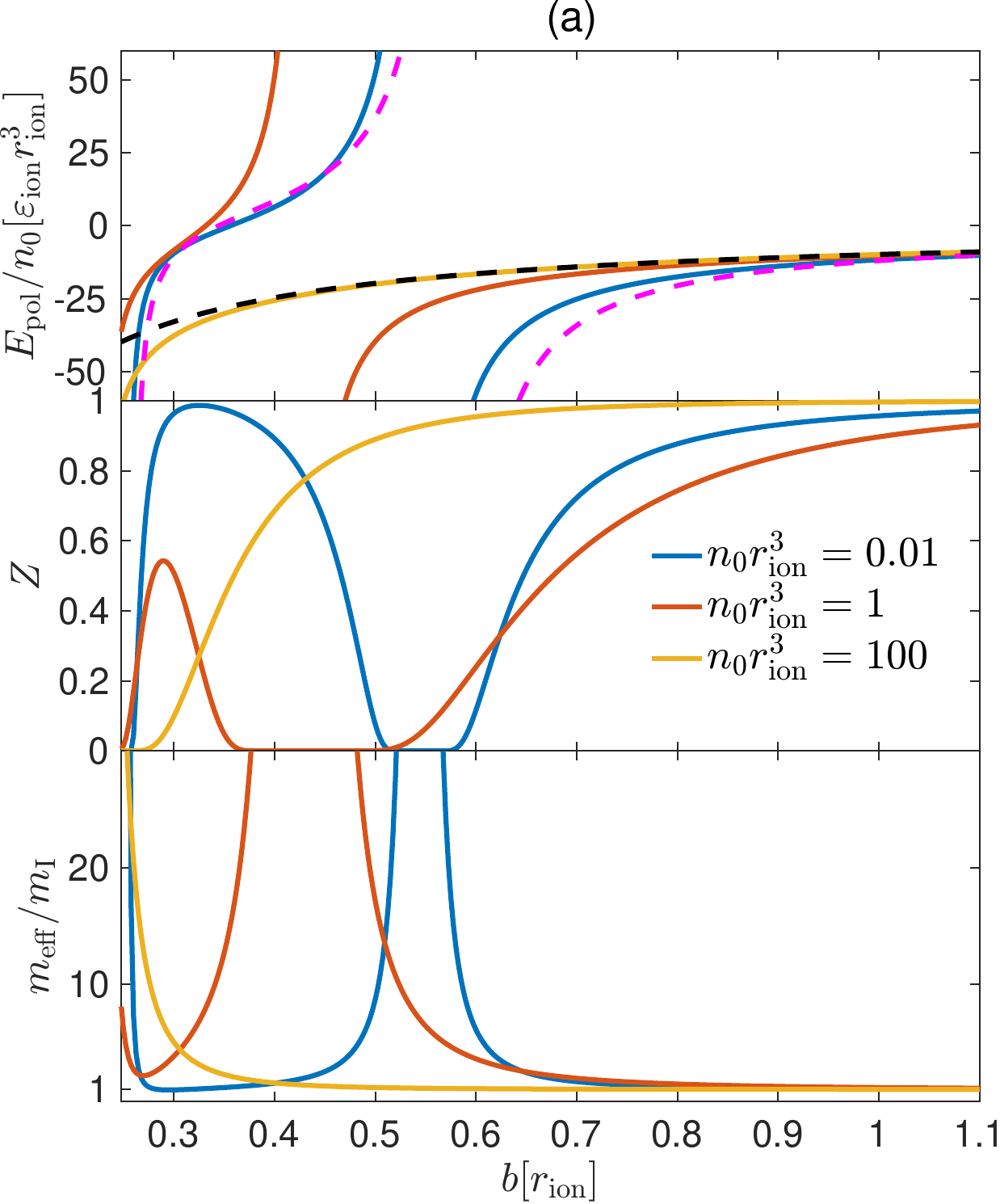}
    \end{subfigure}
    \begin{subfigure}{0.44\textwidth}
        \includegraphics[width = \textwidth,keepaspectratio]{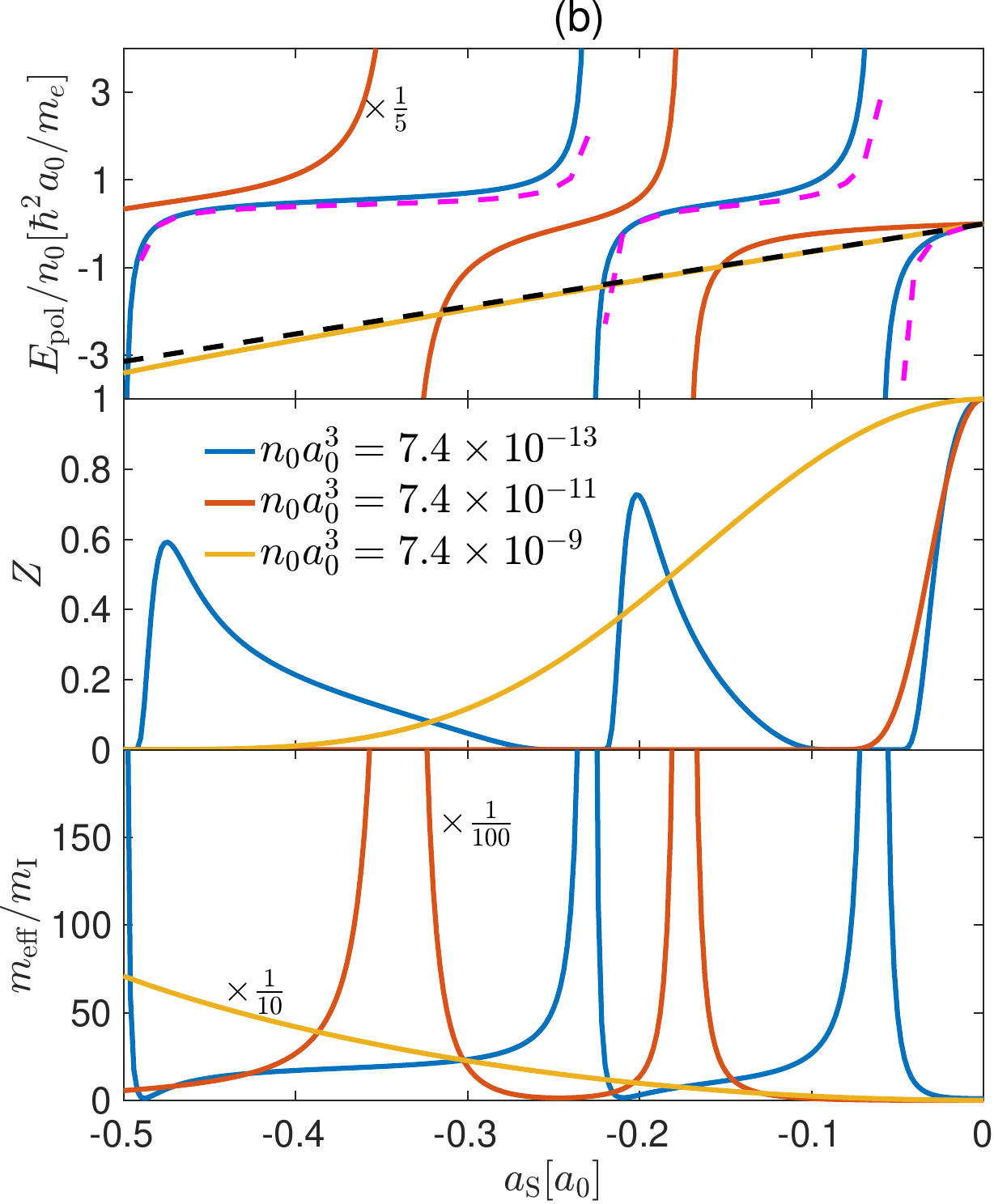}
    \end{subfigure}
    \caption{\justifying The quasiparticle properties of \textbf{(a)} ionic, and  \textbf{(b)} Rydberg impurities across various condensate density regimes as a function of the potential parameters $b$ and $a_{\mathrm{s}}$, respectively. The intraspecies interactions are set to $a_{\mathrm{BB}} = 0.05r_{\mathrm{ion}}$ and $a_{\mathrm{BB}} = 100a_0$ for the ion and Rydberg impurities, respectively.  The dashed magenta (black) lines represents $E_{\mathrm{zr}}$ ($E_{\mathrm{MF}}$). The ratios adjacent to curves represent the multiplication factors of the actual values.
    }
    \label{fig:various_n0}
\end{figure*}

We show the quasiparticle properties of the Rydberg impurity in Fig.\ref{fig:Rydberg_Polaron_Properties_Figure}, computed by the GP (blue lines) and CS (red lines) methods, for $n_0 = 1.48\times 10^{-12}$ ($n_0 = 10^{13} \text{cm}^{-3}$ or $ r_{\mathrm{eff} } n_0^{1/3}\simeq 0.5$), along with the MF results $E_{\mathrm{zr}} = 2\pi a_{\mathrm{IB}} n_0 \frac{m_e}{m_{\mathrm{r}}}$ and $E_{\mathrm{MF}} = \mathcal{U}_{\mathrm{Ryd}}(0)n_0$. We observe all the same generic features of the short-range and ionic impurity in this case. 
For the weak interactions without any available bound states ${a}_\mathrm{s}> -0.05$, the attractive polaron emerges with $E_{\mathrm{pol}} \rightarrow 0$, as well as $Z \rightarrow 1$ when ${a}_\mathrm{s}\rightarrow 0$. Hereafter, the polaron properties strictly deviate from $E_{\mathrm{MF}}$, following the effective-zero range solution $E_{\mathrm{zr}}$. Both repulsive and attractive polarons exist and are sequentially introduced as a new bound state becomes available. Well-defined effective masses accompany the finite quasiparticle weights with descending peaks, and their behavior once again closely follows the key coupling parameter $|a_{\mathrm{IB}}| n_0^{1/3}$. 

Differences in the two-body physics and functional form of the interaction potential manifest in the detailed behavior seen in Fig. \ref{fig:Ionic_Polaron_Properties_Figure} and \ref{fig:Rydberg_Polaron_Properties_Figure}. 
For example, the ranges of the attractive polaron branches in the presence of the bound states are much narrower for the Rydberg potential, compared to those of the ionic impurity. This leads to much sharper increases (decreases) in the effective mass (quasiparticle weight) near the next bound state location. 

\section{\label{sec: Effect of Condensate Particle Interactions} Effects of Bath Particle Interactions and Condensate Density}
We now present a systematic assessment of the quasiparticle nature of impurities across various length scale regimes, including the interparticle distance $n_0^{-1/3}$, the interaction range $r_{\mathrm{eff}}$, and the coherence length $\xi$. 
The effect of intra-species interactions on the quasiparticle nature of long-range impurities has remained relatively unexplored in the literature, particularly for Rydberg impurities, where the functional determinant approach has been employed  \cite{2018_PRA_Schmidt_Rydberg_Polaron_Theory,2024_PRR_Aileen_Rydberg_phenomenology}, which assumes a non-interacting bath. 

In Fig.~\ref{fig:various_n0}, we present the quasiparticle properties of both ionic and Rydberg impurities across a range of condensate densities for a constant finite value of $a_{\mathrm{BB}}$. When $n_0^{1/3} r_\mathrm{eff} \ll 1$, the long-range impurity is rendered effectively short-ranged. As a result, universal features associated with short-range impurities emerge. The polaron energy (blue solid lines) approaches the zero-range MF energy $E_{\mathrm{zr}}$ and, away from scattering resonances, the quasiparticle weight remains close to one. As we move to the intermediate filling regime, $n_0^{1/3} r_\mathrm{eff} \simeq 1$ (red solid lines), deviations from the zero-range results become apparent.  The quasiparticle weights between the scattering resonances are significantly reduced, but remain finite as long as $|a_{\mathrm{IB}}|n_0^{1/3}  \lesssim 1$. In the high filling parameter regime $n_0^{1/3} r_\mathrm{eff} \gg 1$ (yellow solid lines), however, the quasiparticle weight diminishes to arbitrarily small values, even between the scattering resonances. Therefore, the quasiparticle description breaks down when the impurity volume is filled by a large number of bath particles whenever the interaction potential supports at least one bound state. 
When the latter condition is not held, the attractive polaron retains a finite quasiparticle weight independent of density. 
\begin{figure*}[!htb]
    \begin{subfigure}{0.33\textwidth}
        \includegraphics[width = \textwidth]{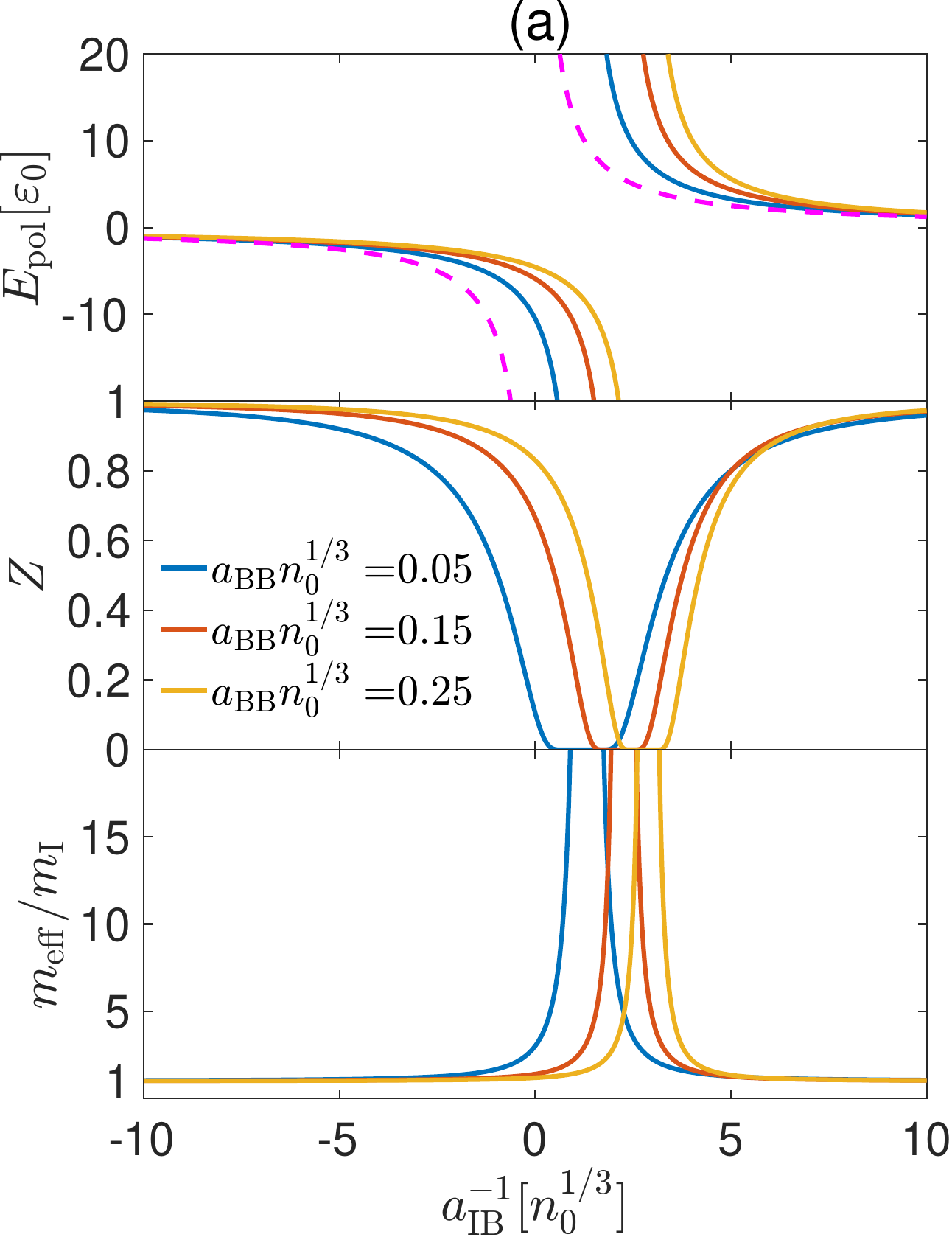}
    \end{subfigure}
    \begin{subfigure}{0.325\textwidth}
        \includegraphics[width = \textwidth]{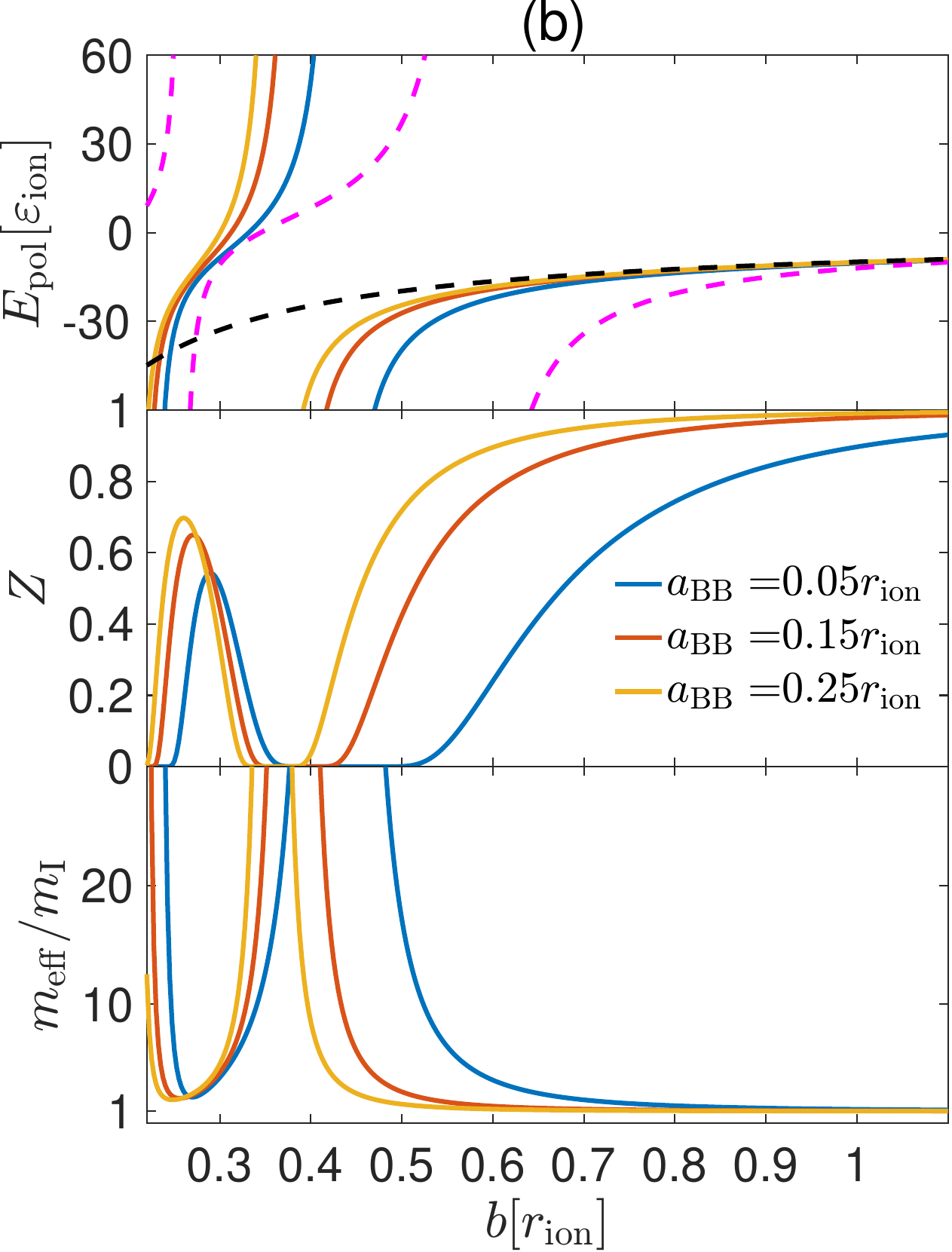}
    \end{subfigure}
    \begin{subfigure}{0.33\textwidth}
        \includegraphics[width = \textwidth]{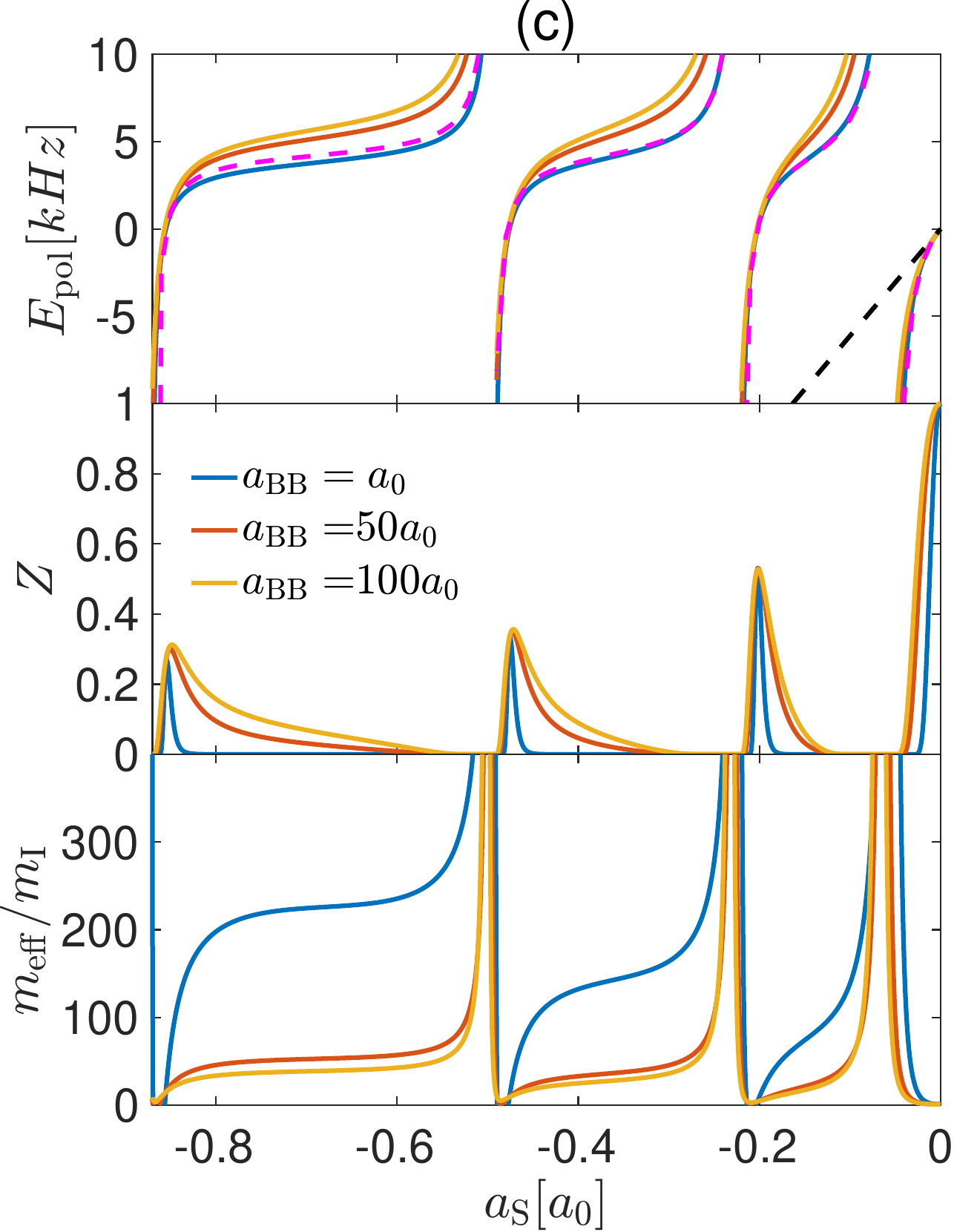}
    \end{subfigure}
    \caption{\justifying The quasiparticle properties of \textbf{(a)} contact, \textbf{(b)} ionic, and \textbf{(c)} Rydberg impurities for various bath-bath scattering length $a_{\mathrm{BB}}$. The condensate density is set to ${n}_0 r_{\mathrm{ion}}^3=1$ for ionic, and $n_0a_0^3 = 1.48 \times 10^{-12}$ ($n_0^{1/3}r_{\mathrm{eff}} \simeq 0.5$) 
    for Rydberg impurities. The dashed magenta (black) lines represents $E_{\mathrm{zr}}$ ($E_{\mathrm{MF}}$). }
    \label{fig:various_aBB}
\end{figure*}

In the weak-coupling regime, the effective mass of long-range impurities converges to unity, irrespective of the bath density, provided that no bound state is supported. This behavior is similar to the case of zero-range impurities. However, we observe that when a bound state is supported, the effective mass at the weak coupling limit $a_{\mathrm{IB}}\to 0$, significantly depends on the filling parameter $r_{\mathrm{eff}}n_0^{1/3}$. In the dilute-density limit, the effective mass remains close to unity in the weak-coupling regime, however, it increases with the increasing filling parameter. This behavior may be attributed to the occupation of molecular states when such states become accessible. In general, a stronger coupling parameter indicates a greater effective mass for any impurity type. 

In the presence of intraspecies interactions, an additional distinct behavior emerges across the transition from low to high filling parameters: each scattering resonance shifts toward deeper impurity-bath interaction. As seen in Fig.\ref{fig:various_n0}(a) and (b), the polaron energy curves, coincident with $E_{\mathrm{zr}}$ at the low density, shift and eventually converge to $E_{\mathrm{MF}}$ at the high density. We explain this shift by an effective modification of the impurity-bath potential due to the presence of the repulsive condensate field, as described in Ref.\cite{2022_SciPost_Schmidt_GP_Bose_Polaron, 2018_PRA_Trapping_Collapse,2021_PRL_Perish_Screening_Short_Range_Impurity} for the case of short-range impurity. This happens when the coherence length of the condensate $\xi \propto (a_{\mathrm{BB}}n_0)^{-1/2}$ becomes relevant to the other length scales in the problem. If the condensate can support density modifications within the interaction range of the impurity, i.e. $\xi \lesssim r_{\mathrm{eff}}$, the bare impurity potential is effectively revised: $U_{\mathrm{eff}} (\mathbf{r}) = U_{\mathrm{IB}}(\mathbf{r}) + g_{\mathrm{BB}} |\Psir|^2$ \cite{2022_SciPost_Schmidt_GP_Bose_Polaron}. Subsequently, the effective potential begins to support bound states at interaction strengths deeper than those supported by the bare potential alone. As the first scattering resonance shifts far deeper, the $E_{\mathrm{MF}}$ curve provides a reliable guideline over a region in the parameter space, noticeably broader than that observed at low densities.

Similar qualitative trends can also be probed by tuning the coherence length through the variation of intraspecies scattering length $a_{\mathrm{BB}}$. In Fig.\ref{fig:various_aBB}, the quasiparticle properties are plotted for three $a_{\mathrm{BB}}$ values of each impurity type with a fixed density. For the short-range impurity, the transition between the repulsive and attractive polaron branches shifts towards smaller ${a}_{\mathrm{IB}}$ values, and the quasiparticle weight approaches unity more rapidly as $a_{\mathrm{BB}}$ increases. For the long-range impurities, the next bound state emerges for deeper potentials, and the fraction of spectral weight lost to the molecular states is reduced. Stronger $a_{\mathrm{BB}}$ results in higher quasiparticle weights for any type of impurity. This phenomenon can be intuitively understood by considering the reduced compressibility of the Bose gas as $a_{\mathrm{BB}}$ becomes larger. The condensate's resistance against the external pressure in the presence of the impurity increases, resulting in a higher likelihood of preserving the initial many-body state. By the same reasoning, the effective mass of the impurity decreases considerably. A compelling question to investigate is the extent to which stronger bath-bath interactions can further sustain these enhanced quasiparticle weights.

\section{\label{sec: Conclusion} DISCUSSION and CONCLUSION }

The reported results can be investigated in ultracold gas experiments. Radio-frequency spectroscopy \cite{2016_PRL_Bruun_Bose_Polaron_Experiment,2016_PRL_Cornell_Bose_Polaron_Experiment,2018_PRL_Schmidt_Rydberg_polaron_experiment}, Ramsey contrast measurements \cite{2024_Arxiv_Cambridge_Bose_Polaron}, and oscillatory perturbations to the impurity via external drives \cite{2012_PRA_1D_Bose_Polaron_Experiment} have been employed to obtain the polaron energy, quasiparticle weight, and effective mass, respectively. Although, in principle, the interaction between an ionic impurity and bath particles can be tuned through a Feshbach resonance \cite{weckesser2021observation}, no analogous tuning mechanism exists for the scattering length $a_{\mathrm{s}}$ in the case of a Rydberg impurity. Instead, variation of the Rydberg atom’s principal quantum number may enable access to distinct regimes with a different number of available bound states \cite{2018_PRL_Schmidt_Rydberg_polaron_experiment,2024_PRR_Aileen_Rydberg_phenomenology}.

We employed two well-established theoretical methods to assess and compute the quasiparticle properties of long-range impurities in a weakly-interacting Bose condensate. We additionally obtained new analytical expressions \eqref{CS_effective_mass_contact} and \eqref{GP_effective_mass_contact} for the effective mass of the short-range impurity, going beyond Fr\"ohlich interactions in the Hamiltonian.  

We considered how polaron properties are affected by the interplay between the three different length scales of the system. We found that the coupling parameter $|a_{\mathrm{IB}}|n_0^{1/3}$, determined by the zero-energy scattering length of the long-ranged impurity, remains the key parameter to assess the quasiaparticle nature of the impurity. In the weak $|a_{\mathrm{IB}}| n_0^{1/3} \ll 1$ and intermediate $|a_{\mathrm{IB}}| n_0^{1/3} \simeq 1$ coupling parameter regimes, the impurity can always be associated with a finite quasiparticle weight. However, this finite weight is significantly suppressed as the impurity volume is increasingly occupied by particles, which can be characterized by the filling parameter $r_{\mathrm{eff}} n_0^{1/3}$. Eventually, as the filling parameter becomes large, the quasiparticle weights become arbitrarily small. We also found that the coherence length $\xi$ becomes a significant factor in determining the polaron properties when $\xi \lesssim r_{\mathrm{eff}}$, as the condensate field begins to effectively modify the bare impurity–bath interaction potential. This leads to a shift of resonance positions. Finally, irrespective of the impurity type or density regime, the stronger intraspecies interactions enhance the quasiparticle weight and reduce the effective mass. 

Future research should address the following issues. First, the saddle point analysis of the polaron properties between two scattering resonances does not account for the partial spectral weight shared with the molarons. A more rigorous theoretical framework is required to describe the coexistence of both atomic and molecular polarons ("molarons") in this regime. The GP theory can be extended to incorporate the molecular impurity field created by the atomic impurities after the quench. Secondly, the regime $\xi \lesssim r_{\mathrm{eff}}$ warrants further investigation. As $\xi$ decreases, the spectrum may exhibit significant shifts in positions of resonances, and sharper features in the polaron branches, away from resonances.

\acknowledgments
We would like to express our gratitude to Aileen A. T. Durst for her valuable contributions and insightful discussions throughout the development of this project.

\bibliography{refs}

\appendix
\section{\label{Appendix:Time-dependent and Saddle Point}Time-dependent results and saddle-point solution}
The conventional approach used to investigate the behavior and properties of the impurity-bath system involves the instantaneous introduction of the impurity at time $t = 0$.
Effectively, this means quenching the interaction strength from zero to a finite value. 
The autocorrelation function or Loschmidt echo, $S(t) = \braket{\Psi(t)}{\Psi(0)}$, provides abundant information regarding the polaron. It quantifies the degree to which the interacting state remains similar to the initial non-interacting state, and its asymptotic behavior is interpreted as the quasiparticle weight or residue, $Z = \lim_{t\rightarrow \infty} S(t)$. By Fourier transforming $S(t)$ to obtain the absorption spectrum, one can also gain information about the polaron energy and broadening. 

Here we discuss the physical relevance of the saddle point solutions to the polaron properties by analyzing the time evolution of the autocorrelation function $|S(t)|$ and the coherent-state amplitudes $\betak(t)$ in Eq.\,\eqref{time_dependent_betak_eqns} for attractive and repulsive interactions. 
The time evolution of $\betak(t)$ is computed by solving the coupled set of first-order differential equations $\partial_t \vec{\betak}= \mathbf{A}\vec{\betak} + \mathbf{f}$ obtained from the matrix form of Eq.\,\eqref{time_dependent_betak_eqns} \cite{2016_PRL_Demler_Coherent_State_Beyond_Frohlich,2020_PRL_Schmidt_Dynamical_Variational_Approach,2021_PRL_Bruun_Ionic_Polaron}. The Loschmidt echo is then numerically evaluated following $S(t)  = \exp{-i\phi(t) -\frac{1}{2} \sumk{k} |\betak(t)|^2}$. Here, the time evolution is studied to show the convergence to the saddle point behavior. 

\begin{figure}
    \includegraphics[width = 1\columnwidth]{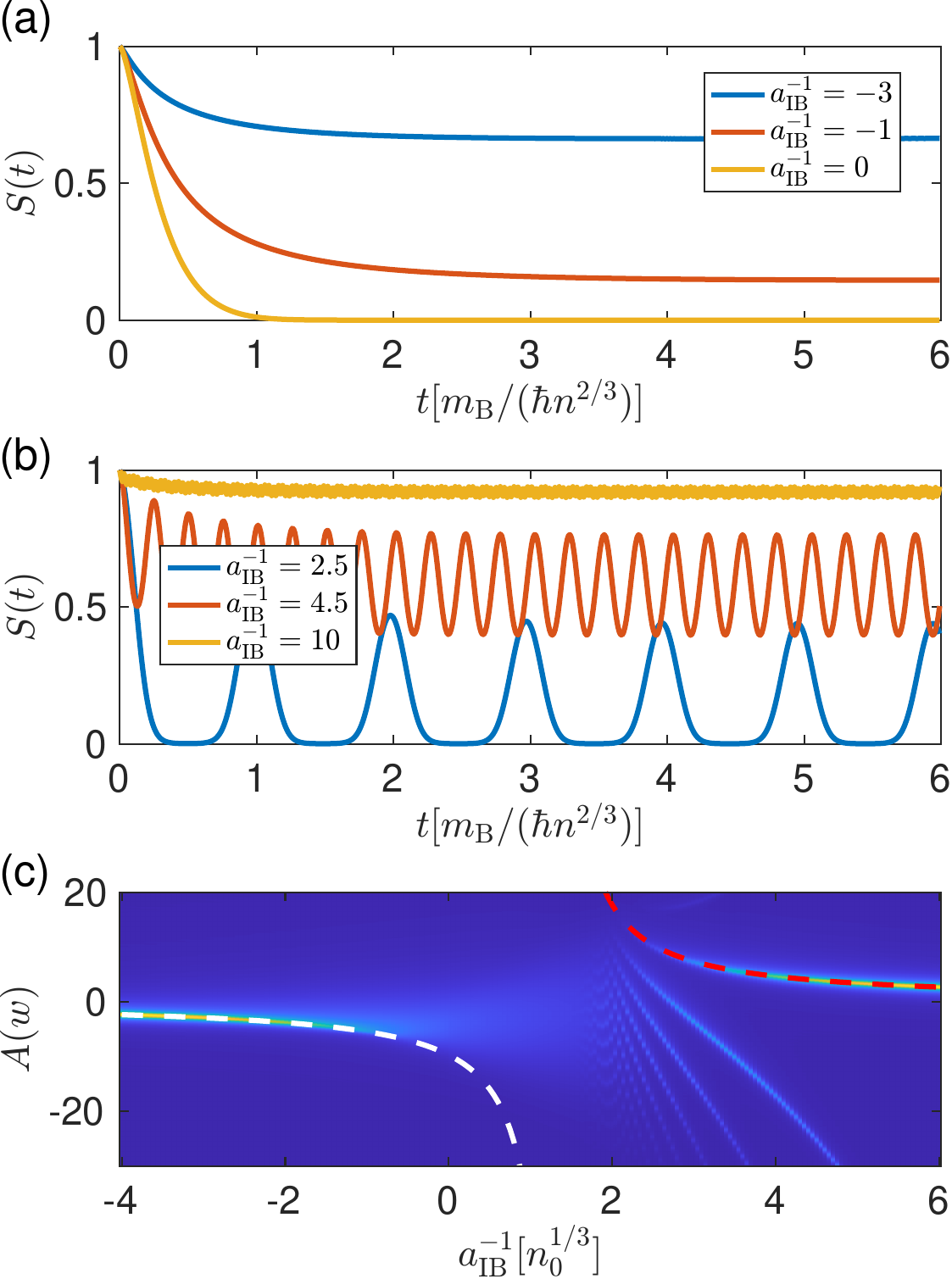}
    \caption{\justifying The time evolution of the autocorrelation function $|S(t)|$ of \textbf{(a)} Attractive, and \textbf{(b)} Repulsive contact polarons, for various impurity-bath scattering lengths $a_{\mathrm{IB}}$. \textbf{(c)} The absorption spectrum of the impurity $A(\omega)$. The dashed white (red) lines are the saddle point attractive (repulsive) polaron energy solutions, computed by \eqref{contact_polaron_energy_CS}. \label{fig:Overlap_function}}
\end{figure}

The saddle point solution admits a clear physical interpretation in the context of the stable attractive polaron. As shown in Fig. \ref{fig:Overlap_function}(a), the overlap with the initial non-interacting state converges to a finite value defining the quasiparticle weight of the particle. The weight is significantly depressed as $a_{\mathrm{IB}}$ approaches the scattering resonance, due to its incoherent distribution into the scattering continuum at higher frequencies  \cite{2016_PRL_Demler_Coherent_State_Beyond_Frohlich} (See Fig. \ref{fig:Overlap_function}(c)). 

A remarkable difference arises in the repulsive branch: The steady-state behavior is oscillatory, and its mean value (amplitude) becomes greater (smaller) for weaker $a_{\mathrm{IB}}$, as illustrated in Fig. \ref{fig:Overlap_function}(b). The oscillations are attributed to the presence of the bound states, therefore, the loss of the spectral weight is coherent, as observed in the absorption spectrum  $A(\omega) = 2 Re \int_0^{\infty} dt e^{i\omega t} S(t)$ shown in the Fig. \ref{fig:Overlap_function}(c). Due to this oscillatory behavior, the interpretation of the saddle point solution, which computes the mean value of the oscillations, needs clarification for the repulsive branch. In the weak coupling limit, where the mean value is finite and oscillation amplitudes become negligible, as in the orange line in Fig. \ref{fig:Overlap_function}(b), it physically corresponds to the coherent amplitudes of the long-living repulsive polaron. For strong couplings, however, the bound states are heavily populated, leading to a significant redistribution of spectral weight toward the so-called "molaron" states. But the polaron properties, such as the effective mass and the energy, derived from the saddle point solutions of the beyond Fr\"ohlich model, can not capture these molaron branches in the absorption spectrum. Yet, they provide a good quantitative agreement in the weak and intermediate coupling regimes and may offer a fair qualitative description for strong $a_{\mathrm{IB}}$. This agreement is clearly seen in the absorption spectrum Fig.\ref{fig:Overlap_function}(c), where the red dashed line, computed by the steady state solutions \eqref{contact_polaron_energy_CS}, closely follows the repulsive polaron branch in the absorption spectrum. 
\section{\label{Appendix:Numerics} Numerical implementation}
We solve Eq.\,(\ref{time_independent_betak_eqns}) numerically on a grid of momenta. In this representation, Eq.\,(\ref{time_independent_betak_eqns}) can be written in the matrix form $\mathbf{A} \Vec{\beta} = \mathbf{v}$, where $\betaVec$ is a vector consisting of $\betak$ variables, discretized in accordance with the symmetry of the problem. 
For quasiparticle properties calculated in the ground state of the impurity-bath system, such as the polaron energy and residue, $\betak$ values depend only on $|\mathbf{k}|$, i.e. $\betakk{k} = \beta_k$. Therefore, $\Vec{\beta} = (\beta_{k_1}, \beta_{k_2}, .., \beta_{k_K})^T$, and $k_l = l dk$, $Kdk = \Lambda$, where $\Lambda$ is the momentum cutoff, $dk$ is the step size in momentum magnitude, and $l$ is a positive integer up to $K$. The $\mathbf{A}$ matrix is a $K$-by-$K$ matrix, whose elements $A_{lm}$ contain the coefficients in linear equations \eqref{symmetric_time_independent_betak_eqns}. But the degeneracy of $\betak$ variables corresponding to the same $\beta_k$ must be regarded carefully. Consider the summation over $\mathbf{k'}$ vectors in \eqref{symmetric_time_independent_betak_eqns} for a given $\mathbf{k}$. One can assume $\mathbf{k}$ is along $z$-axis without loss of generality, then the summation is azimuthally symmetric: 
\begin{eqnarray}
    &&\frac{W_{\mathbf{k}}}{V} \sumk{k'}  \UIBk{k-k'}  W_{\mathbf{k'}}  \betakk{k'} = \\
    &&\frac{W_k}{(2\pi)^2} \int_0^{\Lambda} dk' k'^{2} W_{k'} \beta_{k'} \int_0^{\pi} d\thetak{k'} \sin{\thetak{k'}} \UIBk{|k-k'|},\nonumber 
\end{eqnarray}
where $|\mathbf{k-k'}| = \sqrt{k^2 + k'^2 -2kk'\cos{\thetak{k'}}}$. Define the polar integration with a function $f_{PI}(k,k') \equiv \int_0^{\pi} d\thetak{k'} \sin{\thetak{k'}} \UIBk{|k-k'|}$. Then, Eq.\eqref{symmetric_time_independent_betak_eqns} becomes:
\begin{eqnarray}
    - \frac{\sqrt{N}}{V} \mathcal{U}_{\mathrm{IB}}(k) W_{k}&=& \Omega_k \beta_k  \\
     &+& \frac{W_k}{(2\pi)^2} \int_0^{\Lambda} dk' k'^{2} W_{k'} f_{PI}(k,k') \beta_{k'}  \nonumber
\end{eqnarray}
The elements of the matrix $\mathbf{A}$ and the vector $\mathbf{v}$ in $\mathbf{A} \Vec{\beta} = \mathbf{v}$ are given by:
\begin{eqnarray}\label{A_matrix_elements_1d}
    A_{ll} = \Omega_{k_l} + \frac{W_{k_l}^{2}}{(2\pi)^2} dk \  k_l^2 f_{PI}(k_l,k_l), \nonumber \\
    A_{ll'} = \frac{W_{k_l}W_{k_{l'}}}{(2\pi)^2} dk \  k_m^2 f_{PI}(k_l,k_{l'}),\nonumber \\
    v_l = - \sqrt{n_0} \left(\frac{dk}{2\pi}\right)^{3/2} \mathcal{U}_{\mathrm{IB}}(k_l) W_{k_l},
\end{eqnarray} 
where $V = (2\pi/dk)^3$ is used for $v_l$ coefficients. The solution for $\beta_k$ variables can be found by a matrix inversion: $\Vec{\beta} = \mathbf{A}^{-1} \mathbf{v}$.

Similarly, the matrix and vector elements for the numerical solution of $\delpsikk{k0}$ variables in the set of linear equations \eqref{psikr_equations} are:
\begin{eqnarray}\label{A_matrix_elements_1d_GP}
    A_{ll} = 1 + \frac{1}{4\pi^2}dk \ k_l^2 \mathcal{W}_{k_l} f_{PI}(k_l,k_l), \nonumber \\
    A_{ll'} = \frac{1}{4\pi^2}dk \ k_{l'}^2 \mathcal{W}_{k_l} f_{PI}(k_l,k_{l'}),\nonumber \\
    v_l = - \sqrt{n_0} \left(\frac{dk}{2\pi}\right)^{3/2} \mathcal{U}_{\mathrm{IB}}(k_l) \mathcal{W}_{k_l},
\end{eqnarray} 
where $\mathcal{W}_k \equiv \frac{\epsilonkr}{\Ekr^2}$. Note that each term in Eq.\eqref{psikr_equations} is divided by $\Ekr^2$.

Both the momentum step size $dk$ and the cutoff momentum $\Lambda$ are refined until convergence of the results is achieved, with the parameter set reported in the text corresponding to the converged values. For the Rydberg impurity, the optimization of these parameters is more demanding, as the length scales determined by the internodal distance of the Rydberg potential and the overall system size differ by several orders of magnitude. In this case, $dk$ and $\Lambda$ are chosen such that the fine structure of the potential is accurately resolved while simultaneously ensuring convergence with respect to the total system size.
\section{\label{Appendix:Effective Mass Equations} Low-velocity expansion of the polaron energy}
We calculate the effective mass of the impurity by analyzing the low-velocity corrections to ground-state polaron energy. To this end, we replace $\betak$ in Eq.\,\eqref{time_independent_betak_eqns} by its expansion \eqref{betak_low_velocity_expansion} around the zero-velocity solution. This leads to equations of the zeroth, first, and second order in the velocity $\mathbf{V}$, expressed in terms of the zero-velocity solution $\betakk{k0}$, and the corresponding order corrections $\alpha_k$ and $\gamma_k$: 

\begin{eqnarray} \label{CS_order_eqns}
    0 &=& \Omega_{\mathbf{k0}} \betakk{k0} + \frac{\sqrt{N}}{\mathcal{V}} \UIBk{k} W_{\mathbf{k}} + \frac{1}{\mathcal{V}} \sumk{k'}  \hkkPrime  \betakk{k'0} \\
    0 &=& \Omega_{\mathbf{k0}} \betakk{k0} \alpha_k \kdotv{k} - \betakk{k0} \kdotv{k} \nonumber \\
    && \ \ \ \ \ \ \ \ \ \ + \frac{1}{\mathcal{V}} \sumk{k'} \hkkPrime \betakk{k'0} \alpha_{k'} \kdotv{k'}  \\   
    0 &=& - \alpha_k \betakk{k0} (\kdotv{k})^2 + \Omega_{\mathbf{k0}} \betakk{k0} \gamma_k(\kdotv{k})^2 \nonumber\\
    && \ \ \ \ \ \ \ \ \ \ +\frac{1}{\mathcal{V}} \sumk{k'} \hkkPrime \betakk{k'0} \gamma_{k'} (\kdotv{k'})^2  
\end{eqnarray}
where $\hkkPrime \equiv \UIBk{k-k'} V_{\mathbf{k k'}}^{(1)} + \UIBk{k+k'} V_{\mathbf{k k'}}^{(2)}$. In case of the short-range impurity, the zero-velocity solution is given by \eqref{betak_solutions_short_range}, and the order corrections can be calculated exactly: 
\begin{eqnarray} \label{short_range_order_corrections}
    \alpha_k &=& \frac{1}{\Omega_{k0}}, \\
    \gamma_k &=& \frac{1}{\Omega_{k0}^2}\frac{1}{1 + \frac{g_{\mathrm{IB}} }{\mathcal{V}} \sumk{k'} \frac{W_{k'}^2}{\Omega_{k'0}}}.
\end{eqnarray}
For the long-range impurities, analytical expressions for the exact solutions are infeasible. Instead, one can approximate the order corrections as $\alpha_k = \frac{1}{\Omega_{k0}}$ and $\gamma_k = \frac{1}{\Omega_{k0}^2}$, leading only to minor quantitative changes in the effective mass compared to the exact numerical solution. The zeroth order solution $\betakk{k0}$ can be found by Eq.\,\eqref{symmetric_time_independent_betak_eqns}. 

In the case of the perturbative GP theory, we derive a set of equations for the zero-velocity solution $\delpsikk{k0}$, and the order corrections $\alpha_k$ and $\gamma_k$, similar to \eqref{CS_order_eqns}:
\begin{widetext}
\begin{eqnarray} \label{GP_order_eqns}
    \frac{\phi_0}{\sqrt{\mathcal{V}}} \UIBk{k} \epsilonkr  &=& -\Ekr^2 \delpsikk{k0} +  \frac{1}{\mathcal{V}} \sumk{k'}  \hkkPrime  \delpsikk{k'0} \\
    \frac{\phi_0}{\sqrt{\mathcal{V}}} \UIBk{k} \kdotv{k} &=& -\Ekr^2 \alpha_k \delpsikk{k0}  \kdotv{k}  
    +  \frac{1}{\mathcal{V}} \sumk{k'} \hkkPrime \delpsikk{k'0} \alpha_{k'} \kdotv{k'} 
    -  \frac{\kdotv{k}}{\mathcal{V}} \sumk{k'} \UIBk{k-k'} \delpsikk{k'0}   \\  
    \Ekr^2 \delpsikk{k0} \gamma_k (\kdotv{k})^2 &=& \delpsikk{k0} (\kdotv{k})^2 
     +\frac{1}{\mathcal{V}} \sumk{k'} \bigg[ \hkkPrime \delpsikk{k'0} \gamma_{k'} (\kdotv{k'})^2 
     - (\kdotv{k}) (\kdotv{k'}) \UIBk{k-k'} \delpsikk{k'0} \alpha_{k'}\bigg]
\end{eqnarray}
\end{widetext}
where $\hkkPrime \equiv g_{\mathrm{BB}}n_0 \UIBk{k+k'} -(\epsilonkr+g_{\mathrm{BB}}n_0)\UIBk{k-k'}$.
We find $\delpsikk{k0}$ by solving Eq.\,\eqref{psikr_equations} with setting $\mathbf{V} = 0$. For the short-range impurity, the zero-velocity solution takes the form \eqref{delpsik_solution_short_range} at $\mathbf{V} = 0$, and the order corrections can be obtained exactly: 
\begin{eqnarray} \label{short_range_order_corrections_2}
    \alpha_k &=& \frac{1}{\epsilonkr}, \\
    \gamma_k &=& \frac{1}{\Ekr^2} \frac{1}{1 + \frac{g_{\mathrm{IB}} }{\mathcal{V}} \sumk{k'} \frac{\epsilon_{k'_r}}{E_{k'_r}^2}}.
\end{eqnarray}
Similar to the CS ansatz, we determine $\delpsikk{k0}$ numerically for the long-range impurities, and approximate the order corrections by $\alpha_k = \frac{1}{\epsilonkr}$ and $\gamma_k = \frac{1}{\Ekr^2}$. 

The effective mass of the short-range impurity, for any mass ratio $m^*$, is given in the CS and GP ansatz, respectively, as
\begin{eqnarray} \label{GP_effective_mass_contact_Appendix}
    \frac{m_{\mathrm{eff}}}{m_{\mathrm{I}}}= 1+ \frac{4 \sqrt{\pi}}{3\sqrt{a_{\mathrm{BB}}m^*(m^*+1)}}  \frac{5/8}{ (\aIBinverse-a_{GP,+}^{-1})^2} , \nonumber \\ 
\end{eqnarray}
\begin{eqnarray}  \label{CS_effective_mass_contact_Appendix}
    \frac{m_{\mathrm{eff}}}{m_{\mathrm{I}}} = 1 &+& \left(\frac{128}{3 m^* } \frac{1}{(\aIBinverse-\aPlusinverse)^2}\right) \nonumber \\
    &\times& \int \frac{dk \ k^2}{\sqrt{k^2 + \tau} \left( k/m^* + \sqrt{k^2 + \tau}\right)^3},
\end{eqnarray}
where $\tau \equiv 16 \pi a_{\mathrm{BB}}$ and the latter integral yields $\frac{2}{15\sqrt{
\tau}}$ when $m^* =1$. 
\section{\label{Appendix:GP Equation Derivation} Alternative derivation of the GP equation for finite impurity at low velocity limit}
An alternative derivation of the GP equation \eqref{GP_Equation} used in the perturbative GP theory is provided here. This derivation highlights the connection between the LLP framework for the GP fields $\Psi(\mathbf{r})$ and the GP equation in \eqref{GP_Equation}, clarifying the implicit assumptions involved.
We start with the total GP energy functional expressed in terms of the condensate field $\Psir$, obtained after the LLP transformation into the co-moving frame of the impurity \cite{1962_Annals_Gross_Motion_of_Impurity,2020_PRR_Fleisheur_Exact_1D_Bose_Polaron,2022_SciPost_Schmidt_GP_Bose_Polaron}
\begin{eqnarray}
    K_{\mathrm{GP}} &=& \frac{\left[ \ptot - \int d^3 \mathbf{r} \Psirconj (-i\hbar \nabla)\Psir\right]^2}{2m_{\mathrm{I}}} \nonumber \\
    &+& \int d^3\mathbf{r} \Bigg\{ \frac{\hbar^2}{2m_{\mathrm{r}}} |\nabla \Psi(\mathbf{r})|^2 + \left[U_{\mathrm{IB}}(\mathbf{r})-\mu \right]|\Psi(\mathbf{r})|^2 \nonumber \\
    &&\ \ \ \ \ \ \ \ \ + \frac{g_{\mathrm{BB}}}{2} |\Psi(\mathbf{r})|^4\Bigg\},
\end{eqnarray}
where the first line corresponds to the real-space counterpart of the momentum-space representation $\frac{(\ptot - \sumk{k} \hbar \mathbf{k} b_{\mathbf{k}}^{\dagger}b_{\mathbf{k}})^2}{2m_{\mathrm{I}}}$ in $\hat{\mathcal{H}}_{\mathrm{LLP}}$ \eqref{LLP_Hamiltonian}. Here the phonon momentum is defined as $\mathbf{p}_{\mathrm{B}} \equiv  \int d^3 \mathbf{r} \Psirconj (-i\hbar \nabla)\Psir$. Note that the motions of impurity and phonons are involved without any waveform argument $(\mathbf{r}-\mathbf{V}t)$ in the impurity-bath potential $U_{\mathrm{IB}}(\mathbf{r})$. We now apply a variational analysis $\frac{\delta K_{\mathrm{GP}}}{\delta \Psi^*} = 0$ to find the time-independent GP equation, which yields
\begin{eqnarray}
    &\mu& \Psir = -\frac{\hbar^2}{2m_{\mathrm{r}}} \nabla^2 \Psi(\mathbf{r}) + U_{\mathrm{IB}}(\mathbf{r})\Psir + g_{\mathrm{BB}}|\Psi(\mathbf{r})|^2 \Psi(\mathbf{r}) \nonumber \\
    &+& \frac{i\hbar \nabla \Psir \cdot \left(\ptot- \int d^3 \mathbf{r'} \Psi^*(\mathbf{r'}) (-i\hbar)\nabla \Psi(\mathbf{r'})\right)}{m_{\mathrm{I}}} ,
\end{eqnarray}
where the second line corresponds to the contributions due to the non-zero total and phonon momentum. In the time evolution of the impurity-bath system, $\ptot$ is a constant of motion, whereas $\mathbf{p}_{\mathrm{B}}$ and the impurity velocity $\mathbf{V} = (\ptot - \mathbf{p}_{\mathrm{B}})/m_{\mathrm{I}}$ can vary. 
When a fixed velocity vector $\mathbf{V}$ is associated with the impurity, this variation is implicitly neglected. Although this assumption may lead to inaccuracies in describing the time evolution of the polaron, it remains justified in a steady-state analysis of regimes in which the polaron attains a nearly or fully constant momentum. This neglect corresponds to the mean-field treatment of the boson-boson interactions induced by the LLP transformation. In this case, applying the perturbative expansion $\Psir = \phi_0 + \delpsir$ yields the same equation \eqref{GP_Equation_delpsi}.
\end{document}